\documentclass[a4paper, 12pt]{article} 
\usepackage[right=1in,left=1in,top=1in,bottom=1in]{geometry}
\usepackage{graphicx} 
\usepackage{braket}
\usepackage{mathrsfs}
\usepackage[T1]{fontenc} 
\usepackage{amsmath}
\usepackage{pxfonts}
\usepackage[T1]{fontenc}
\usepackage{amssymb}
\usepackage{natbib}
\usepackage{epstopdf}
\usepackage{setspace}
\usepackage{enumitem}
\usepackage[colorinlistoftodos]{todonotes}
\usepackage{sectsty}
\usepackage{wrapfig} 
\sectionfont{\large}
\bibpunct{(}{)}{,}{a}{}{,}
\usepackage{tcolorbox}
\usepackage{subfig}

\usepackage{tikz}   
\usepackage{tikz-3dplot} 

\usepackage[protrusion=true,expansion=true]{microtype} 

\newcommand{\qed}{\nobreak \ifvmode \relax \else
      \ifdim\lastskip<1.5em \hskip-\lastskip
      \hskip1.5em plus0em minus0.5em \fi \nobreak
      \vrule height0.75em width0.5em depth0.25em\fi}

\usepackage{bbm}
\usepackage{MnSymbol}
\usepackage[all]{xy}

\usepackage{xcolor,colortbl,array,amssymb}

   \usepackage{braket}
   \usepackage{amssymb}
   \usepackage{epigraph}
\makeatletter

\title{\Large{The Past Hypothesis and the Nature of Physical Laws } }  

\author{Eddy Keming Chen\thanks{Department of Philosophy,  University of California, San Diego, 9500 Gilman Dr, La Jolla, CA 92093-0119. Website: www.eddykemingchen.net. Email: eddykemingchen@ucsd.edu  }}

\date{\today \vspace{15pt} \\ \small{Forthcoming in Barry Loewer,  Eric Winsberg, and Brad Weslake (eds.), \\ \textit{Time's Arrows and the Probability Structure of the World},   Harvard University Press} }

\begin{document}
\bibliographystyle{authordate1}

\maketitle 


\epigraph{Therefore I think it is necessary to add to the physical laws the hypothesis that in the past the universe was more ordered, in the technical sense, than it is today---I think this is the additional statement that is needed to make sense, and to make an understanding of the irreversibility.}{Richard Feynman (1964 Messenger Lectures)}

\begin{abstract}
If the Past Hypothesis underlies various arrows of time, what is the status of the Past Hypothesis? In this paper, I examine the role of the Past Hypothesis in the Boltzmannian account and defend the view that the Past Hypothesis is a candidate fundamental law of nature. Such a view is known to be compatible with Humeanism about laws, but as I argue, it is also supported by a minimal non-Humean ``governing'' conception of laws. Some worries arise from the non-dynamical and time-dependent character of the Past Hypothesis as a boundary condition, the intrinsic vagueness in its specification, and the nature of the initial probability distribution. I show that these worries do not have much force, and in any case, they become less relevant in a new quantum framework for analyzing time's arrows---the Wentaculus. Hence, the view that the Past Hypothesis is a candidate fundamental law should be more widely accepted than it is now. 

\end{abstract}

\hspace*{3,6mm}\textit{Keywords:  time's arrow, counterfactuals, laws of nature, vagueness, objective probabilities, typicality, scientific explanation, Past Hypothesis, Statistical Postulate, Humeanism, non-Humeanism, minimal primitivism, the Mentaculus, the Wentaculus, quantum statistical mechanics, density matrix realism}   

\newpage

\begingroup
\singlespacing
\tableofcontents
\endgroup



\nocite{albert2000time,  price2004origins}

\section{Introduction}

One of the hardest problems in the foundations of physics is the problem of the arrows  of time. If the dynamical laws are  (essentially) time-symmetric, what explains the irreversible phenomena in our  experiences, such as the melting of ice cubes, the decaying of apples, and the mixing of cream in coffee? Macroscopic systems display an entropy gradient in their temporal evolutions: their thermodynamic entropy is lower in the past and higher in the future. But why does  entropy have this temporally asymmetric tendency? Following \cite{goldstein2001boltzmann}, let us distinguish between the two parts of the problem of irreversibility:
\begin{enumerate}
  \item The Easy Part: if a system is not at maximum entropy, why should its entropy tend to be larger at a later time? 
  \item The Hard Part: why should there be an arrow of time in our universe that is governed by fundamental reversible dynamical laws? 
\end{enumerate}
The Easy Part was studied by \cite{boltzmann2012lectures}[1896]. Crucial to Boltzmann's answer is this:

\begin{itemize}
  \item Key to the Easy Part: states of larger entropy occupy much larger volume in the system's phase space than those states of lower entropy. 
\end{itemize}
So far, answering the Easy Part does not require any time-asymmetric postulates.\footnote{Boltzmann's \textit{Stosszahlansatz} (hypothesis of molecular chaos) is often blamed for introducing an illicit time asymmetry. But it is an innocent theoretical postulate if we understand it correctly---as a typicality or probability measure over initial conditions. See \citep[section 5.5]{goldstein2019gibbs}. } Boltzmann's program is primarily focused on  closed subsytems of the universe. But its success leads us to expect that a Boltzmannian account can work at the universal level. If we model the universe as a mechanical system, we expect that typically, the non-equilibrium state of the universe will  evolve towards higher entropy at later times. 

However, why is the entropy lower in the past? That is the Hard Part.  A proposed answer suggests that it has to do with the initial condition of the universe:
\begin{itemize}
  \item Key to the Hard Part: the universe had a special beginning. 
\end{itemize}
We can introduce this as an explicitly time-asymmetric postulate in the theory, by using the Past Hypothesis:

\begin{description}
  \item[Past Hypothesis (PH)] At the initial time of the universe, the microstate of the universe is in a low-entropy macrostate.\footnote{ The Past Hypothesis was originally suggested in \citep{boltzmann2012lectures}[1896] (although he seems to favor another postulate that can be called the \textit{Fluctuation Hypothesis}) and  discussed in \citep{feynman2017character}[1965]. For recent discussions, see \citep{albert2000time}, \citep{goldstein2001boltzmann},  \citep{callender2004measures, sep-time-thermo}, \citep{lebowitz2008time},  \citep{north2011time},  \citep{loewer2016mentaculus}, and \citep{goldstein2019gibbs}. The memorable phrase `Past Hypothesis' was coined by Albert (2000).} 
\end{description}
Given that some microstates are anti-entropic,  it is standard to introduce  a probability distribution over the microstates compatible with the low-entropy macrostate: 

\begin{description}
  \item[Statistical Postulate (SP)] The probability distribution of the initial microstate of the universe is given by the uniform one (according to the natural measure) that is supported on the macrostate of the universe.
\end{description}
However, a detailed probability distribution may be unnecessary. In the typicality framework, we just need to be committed to a typicality measure:
\begin{description}
  \item[Typicality Postulate (TP)] The initial microstate of the universe is typical inside the macrostate of the universe.\footnote{For more on the notion of typicality and its application in statistical mechanics, see \citep{goldstein2012typicality}, \citep{lazarovici2015typicality}, and \citep{wilhelm2019typical}.}  
\end{description}
Unlike SP, TP is compatible with a variety of measures that agree on what  is typical. PH, SP, and TP are  physical postulates that have an empirical status. The answer to the Hard Part of the problem of irreversibility requires PH and either SP or TP. In fact, we also need to assume (an unconditionalized) notion of probability or typicality to answer the Easy Part. I call the answers to the Easy Part and the Hard Part the \textit{Boltzmannian account} of the arrow of time.

How to characterize the initial macrostate of PH remains an open question. We know that the matter distribution is more or less uniform in the early universe, which is contrary to the usual conception of low entropy. However, the initial gravitational degrees of freedom  are in a special state, providing a sense that the total state of the early universe has low entropy. This observation led \cite{penrose1979singularities} to postulate a geometric version of PH called the \textit{Weyl Curvature Hypothesis}: the Weyl curvature vanishes at the initial singularity. The urgent question is, of course, how to understand this in terms of quantum theory or quantum gravity. Some  steps have been taken in the Loop Quantum Cosmology framework by \cite{ashtekar2016initial}. This is compatible with a Boltzmannian account, but the final details will depend on the exact theory of quantum gravity, which is currently absent. 

In the philosophical literature, a number of objections have been raised against the Boltzmannian account. First,  some criticize the answer to the Easy Part: the explanation is too hand-wavy and not completely rigorous \citep{frigg2007field}. Second, some criticize the answer to the Hard Part, such as  that PH is not even false because the entropy of the early universe is not well-defined \citep{earman2006past}, or  that PH is  not sufficient to explain the thermodynamic behaviors of subsystems \citep{winsberg2004can}, or  that it is \textit{ad hoc}, and therefore, not explanatory (Price 2004; Carroll 2010, p.346).  These are responses in the literature, and more work on these issues is certainly welcome. However, my interest here is different. I take the Boltzmannian account as a starting point. My aim is to explore the conceptual and scientific ramifications of accepting  PH and its explanation of time's arrow. 

In this paper, I focus on the connection between PH  and our concept of fundamental laws of nature. What is the status of PH if the Boltzmannian program turns out to be successful? Can PH be accepted as a candidate fundamental law even though it is a boundary condition of the universe? What differences does it make to our concepts of laws, chances, and possibilities? Can PH be completely expressed in mathematical language? What is the relevance of quantum theory to these issues?
I argue for the following theses:

\begin{description}
  \item[Nomic Status] The Past Hypothesis is a candidate fundamental law of nature.\footnote{A \textit{candidate} fundamental law of nature has all it takes to be a fundamental law of nature, but it may not turn out to be the true  law of the actual world if it makes false predictions. For example, Newton's dynamical law $F=ma$ is a candidate fundamental law of nature but it is not the actual fundamental law.  } 
\end{description}

\begin{description}
  \item[Axiomatic Status] The Past Hypothesis is a candidate axiom of the fundamental physical theory.\footnote{The status of a candidate fundamental law of nature and the status of a candidate axiom of the fundamental physical theory may be equivalent. I distinguish the two theses because some people may be happy to accept one but deny the other. They may be reluctant to call PH a fundamental \textit{law}, perhaps due to it being a boundary condition. }
\end{description}

\begin{description}
  \item[Relevance] Whether the Past Hypothesis has nomic status (and/or axiomatic status)  is  relevant to the success of explaining time's arrows, the metaphysical account of laws, the nature of objective probability, and the mathematical expressibility  of fundamental physical theory. 
  \end{description}

Some of these ideas have been defended along Humean lines, but I think we should accept them regardless of whether we think fundamental laws supervene on matter distribution or are part of the fundamental facts of the world. It turns out they are acceptable even on certain non-Humean frameworks.\footnote{This view is, I think, in the same spirit as the suggestion made by \cite{demarest2019mentaculus}.}  My methodology below is  naturalistic and functionalist. \textit{Whatever plays the role of a fundamental law can be a fundamental law.} Whatever that cannot be derived from more fundamental laws \textit{and} plays the right roles in guiding our inferences about the past and the future, underlying various scientific explanations, high-level regularities, our manifest image of influence and control, and so on, is a candidate fundamental law.\footnote{I do not claim that we should reduce laws to these roles; that would be the strategy of  metaphysical functionalism about laws. Rather, I am merely appealing to the methodology in naturalistic metaphysics of science that I think many people accept independently of the issue of the arrows of time. } To borrow a phrase from \cite{loewer2016mentaculus}, PH ``looks, walks and talks'' like a fundamental law. So, we should interpret it as such.  Hence, I disagree with people who think that even if  PH is true and plays all the roles we suggest, it still cannot be a fundamental law---it may just be a special but \textit{contingent} initial condition. I also disagree with people who think that whether or not  PH is a fundamental law makes no substantive differences.\footnote{\cite{MaudlinMWP} \S4 seems to regard PH as an important  boundary condition but does not think of it as a fundamental law. The disagreement is based on a different view about how laws govern that I call Dynamical Law Primitivism, which I discuss  in \S3.4.  \cite{carroll2010eternity} (p.345) suggests that there is no substantive difference between the statements ``the early universe had a low entropy'' and ``it is a law of physics that the early universe had a low entropy.'' Carroll seems to be worried about the distinction between boundary conditions and laws; I discuss this in \S4.1.}

Here is the roadmap. In \S2, I  provide more details of the Boltzmannian account and discuss the modifications to  PH when we move from classical mechanics to quantum mechanics. The variations result in three types of physical theories: the Classical Mentaculus, the Quantum Mentaculus, and the Wentaculus theories. In \S3, I provide positive arguments for the nomic and axiomatic statuses of PH. Some of these have been mentioned in the literature, but it is worth emphasizing and clarifying the exact argumentative structure. I also put forward a novel argument based on considerations about the nature of the quantum state. In \S4, I discuss some apparent obstacles from recognizing PH as a fundamental law. This has to do with its nature as a boundary condition, the status of the Statistical Postulate and the Typicality Postulate, and the intrinsic vagueness in their specifications. I argue that these worries do not have much force even on some non-Humean views, and they become even less worrisome in the Wentaculus theory.

\section{Variations on a Theme from Boltzmann}

Before we get into the philosophical and conceptual issues, let us be more explicit about what the Boltzmannian account is and how to state PH in that account. Although the Boltzmannian account is more or less the same in classical and in quantum theories, the exact form of PH is subtly different. I will exploit this difference in \S4 to dissolve some of the worries about the classical version of PH. Readers familiar with the standard Boltzmannian statistical mechanics can jump to \S2.3, where a new framework called the \textit{Wentaculus} is introduced.

\subsection{The Classical Mentaculus}

Let us start with the Boltzmannian account in classical statistical mechanics and summarize its basic elements from the ``individualistic viewpoint.''\footnote{I follow the discussion in \citep{goldstein2010approachB}. These do not intend to be rigorous axiomatizations of classical statistical mechanics. What is presented here differs in emphasis from \citep{chen2018IPH}, as here I am explicit about the sources of vagueness, which will be discussed in \S4.}  Let us  consider a classical-mechanical system with $N$ particles in a box of volume $\Lambda = [0, L]^3 \subset \mathbb{R}^3$ and a Hamiltonian $H = H(X) = H(\boldsymbol{q_1}, ..., \boldsymbol{q_N}; \boldsymbol{p_1},...,\boldsymbol{p_n})$ that specifies the standard interactions in accord with Newtonian gravitation, Coulomb's law, and other  forces obeyed by the classical system.

\begin{enumerate}
\item Microstate: at any time $t$, the microstate of the system is given by a point in a $6N$-dimensional phase space,
\begin{equation}
X=(\boldsymbol{q_1}, ..., \boldsymbol{q_N}; \boldsymbol{p_1},...,\boldsymbol{p_n}) \in \Gamma_{total} \subseteq \mathbb{R}^{6N},
\end{equation}
where $\Gamma_{total}$ is the total phase space of the system. 

\item Dynamics: the time dependence of $X_t =(\boldsymbol{q_1}(t), ..., \boldsymbol{q_N}(t); \boldsymbol{p_1}(t),...,\boldsymbol{p_n}(t))$ is given by the Hamiltonian equations of motion:
\begin{equation}\label{HE}
\frac{d \boldsymbol{q_i}(t)}{d t} = \frac{\partial H}{\partial \boldsymbol{p_i}} \text{  ,  } \frac{d \boldsymbol{p_i}(t)}{d t} = - \frac{\partial H}{\partial \boldsymbol{q_i}}.
\end{equation}

\item Energy shell: the physically relevant part of the total phase space is the energy shell $\Gamma \subseteq \Gamma_{total}$ defined as:
\begin{equation}
\Gamma = \{X\in \Gamma_{total}: E\leq H(x) \leq E+\delta E\}.
\end{equation}

We only consider microstates in $\Gamma.$

\item Measure: the measure $\mu_V$ is the standard Lebesgue measure on phase space, which is the volume measure on $\mathbb{R}^{6N}$. The Lebesgue measure on a finite volume can be normalized to yield a probability distribution. 

\item Macrostate: with a choice of macro-variables, the energy shell $\Gamma$ can be partitioned into macrostates $\Gamma_{\nu}$: 
\begin{equation}
\Gamma = \bigcup_{\nu} \Gamma_{\nu}.
\end{equation}
A macrostate is composed of microstates that share similar macroscopic features (i.e., similar values of the macro-variables), such as volume, density, and pressure. 

Caveat: the partition of microstates into macrostates is exact only after we stipulate some choices of the parameters for coarse-graining (the size of the cells) and correspondence (between functions on phase space and thermodynamic quantities). We call these \emph{C-parameters}. Without exact choices of the C-parameters, the partition is inexact and the boundaries between  macrostates are vague.\footnote{For more discussions, see \citep{chen2018NV}.} See Figure 1. It is also expected that given the nature of the actual forces, some partitions will be  superior to others in supporting generalizations in the special sciences.  


\begin{figure}%
    \centering
    \subfloat[Vague boundaries]{{\includegraphics[width=8.5cm]{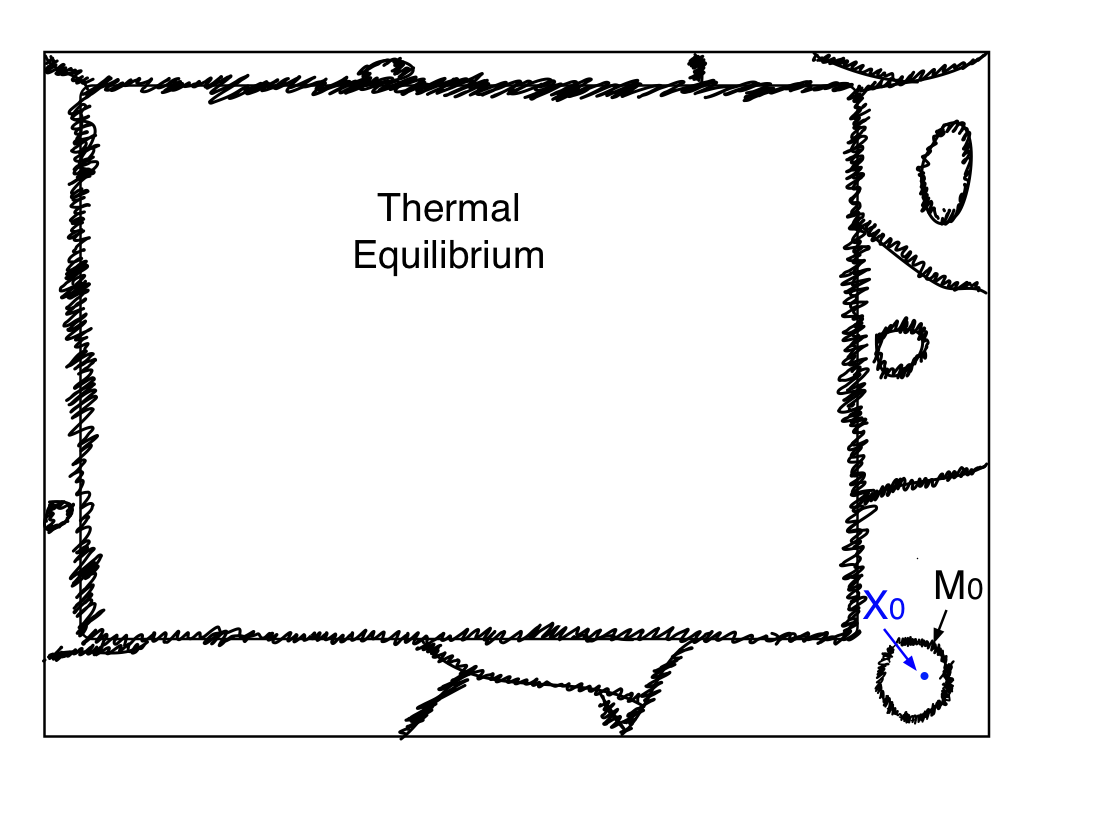} }}%
    \subfloat[Exact boundaries]{{\includegraphics[width=8.5cm]{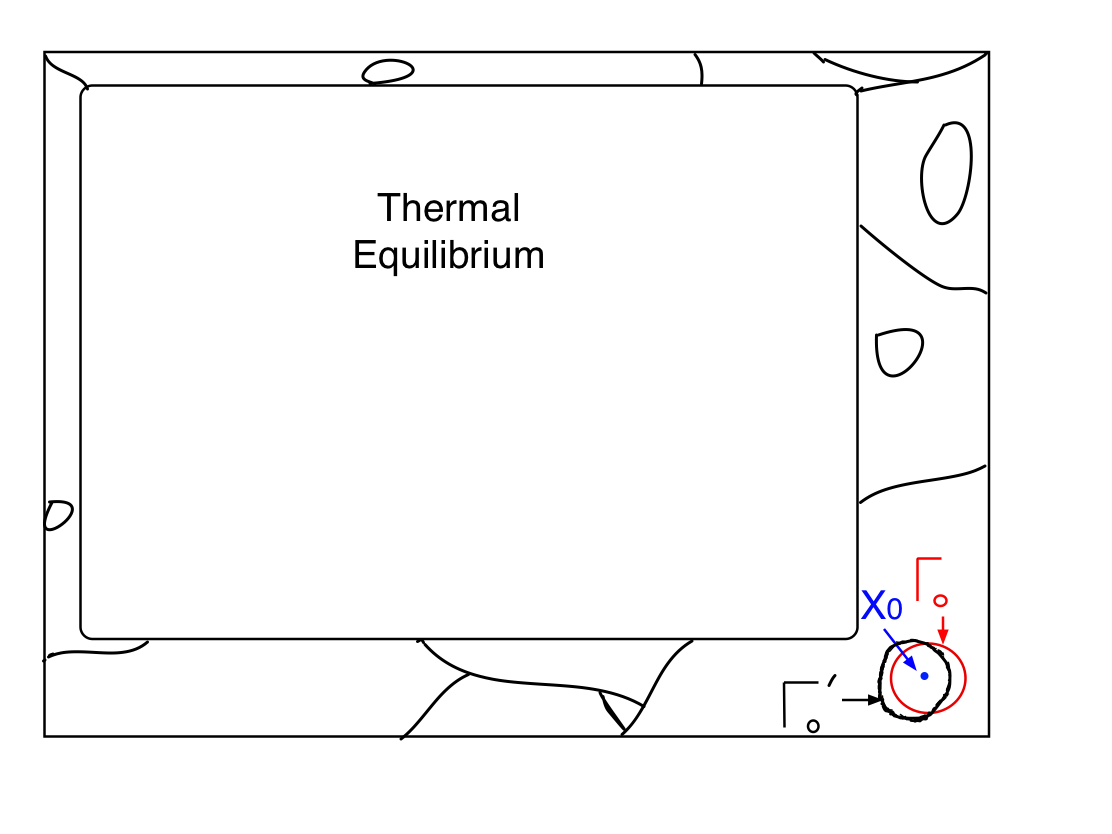} }}%
    \caption{The partition of microstates into macrostates on phase space: (a)  without exact choices of the C-parameters, (b) with exact choices of the C-parameters. $X_0$ represents the actual microstate of the universe at $t_0$. $M_0$ represents the vague boundaries of the PH macrostate. $\Gamma_0$ represents an admissible precisification of $M_0$, where $\Gamma_0'$ represents another admissible precisification. The diagrams are not drawn to scale.}%
    \label{fig:example}%
\end{figure}

\item Unique correspondence: given exact choices of the C-parameters, the macrostates partition the energy shell, and as a consequence, every phase point $X$ belongs to one and only one $\Gamma_{\nu}$. (This point is implied by \#5. But I make it explicit to better contrast it with the situation in quantum statistical mechanics.)

\item Thermal equilibrium: typically, there is a dominant macrostate $\Gamma_{eq}$ that has almost the entire  volume with respect to $\mu_V$:
\begin{equation}
\frac{\mu_V(\Gamma_{eq})}{\mu_V(\Gamma)} \approx 1.
\end{equation}

A system is in thermal equilibrium if its phase point $X\in \Gamma_{eq}.$ 

\item Boltzmann Entropy: the Boltzmann entropy of a classical-mechanical system in microstate $X$ is given by:
\begin{equation}
S_B (X) = k_B \text{log} (\mu_V(\Gamma(X))),
\end{equation}
where $\Gamma(X)$ denotes the macrostate containing $X$. The thermal equilibrium state thus has maximum entropy. 

Caveat: Without exact values of the C-parameters, there will be many admissible choices of the $\Gamma(X)$'s. Moreover, what is admissible is also vague. Since $k_B$ is a scaling constant that plays no direct dynamical role, its value is also vague. Hence, the Boltzmann entropy of a microstate should be understood as a vague quantity. If we stipulate some C-parameters and  the value of $k_B$, we can arrive at an exact boundary for the macrostate that contains $X$ and an exact value of Boltzmann entropy for the system.   

\item Low-Entropy Initial Condition:  on the assumption that we can model the universe as a classical-mechanical system of $N$ point particles, we postulate a special low-entropy boundary condition, which  Albert (2000) calls \emph{the Past Hypothesis} (PH): 
\begin{equation}\label{PH}
 X_{t_0} \in \Gamma_{PH} \text{ , } \mu_V(\Gamma_{PH}) \ll \mu_V(\Gamma_{eq}) \approx \mu_V(\Gamma),
\end{equation}
where $\Gamma_{PH}$ is the PH macrostate with a volume much smaller than that of the equilibrium macrostate.  Hence,  $S_B (X_{t_0})$, the Boltzmann entropy of the  microstate at the boundary, is very small compared to that of thermal equilibrium. Here, $\Gamma_{PH}$ is underspecified; we can add further details to specify the macroscopic profile (temperature, pressure, volume, density) of $\Gamma_{PH}$. 
\end{enumerate}

The answer to the Easy Part of the problem of irreversibility lies in the first eight bullet points, which make plausible the hypothesis of the  typical tendency for a system to evolve to higher entropy towards the future. Even though microstates are ``created equal,'' macrostates are not. Their volumes are disproportionate and uneven. Macrostates with higher entropy have much larger volume in the energy shell. Furthermore, the largest macrostate is by far that of thermal equilibrium. It is plausible that, unless the dynamics are extremely contrived,  a typical microstate starting from a medium-entropy macrostate will find its way through larger and larger macrostates and eventually arrive at thermal equilibrium. That is a process in which a system's entropy gradually increases until it reaches the entropy maximum.\footnote{Boltzmann's original H-theorem \citep{boltzmann2012lectures}[1896] is an attempt to show this. \cite{lanford1975time} produces an exact result for a simple system of hard spheres where the Boltzmann equation is shown to be satisfied for a short duration of time, and hence, Boltzmann entropy is shown to be increasing towards the future. However, it is plausible that the equation continues to be valid and  Boltzmann entropy continues to rise afterwards.} Of course, for the actual universe, there can be exceptions to the entropy increase, such as short-lived fluctuations in which entropy decreases. 

However, this solves only half the problem. If typical microstates compatible with a medium-entropy macrostate will, at most times, increase in entropy towards the future, then typical micorstates compatible with the same macrostate will also, at most times, increase in entropy towards the past. Hence, given the resources so far, we have shown that the medium-entropy macrostate is overwhelmingly likely at an entropy minimum produced by a thermodynamic fluctuation from equilibrium. We are led to the Hard Part of the problem: why is the entropy so much lower in the past direction of time? Enter the  Past Hypothesis. Given PH, the actual microstate starts in an atypical region of the energy shell, in a low-entropy macrostate $M_0$. Suppose we choose a precisification  $\Gamma_0$. Given the Easy Part, typical initial microstates compatible with  $\Gamma_0$ will evolve towards macrostates of higher entropy in the future direction. But there is nothing earlier than $t_0$, as it is stipulated to be the initial time---say, the time of the Big Bang.\footnote{It can also be stipulated that $t_0$ is some time close to the Big Bang, in which case some anti-thermodynamic behavior can be displayed in the short duration before $t_0$.} 

Therefore, if we find the universe to be in a medium-entropy macrostate $\Gamma_t$, say the state we are in right now, then the actual microstate is not like a typical microstate inside $\Gamma_t$, but a special one that is compatible with $\Gamma_0$. The reason that the entropy was lower in the past is because the universe started in a special macrostate, a state of very low Boltzmann entropy. Assuming PH, it is reasonable to expect \textit{with overwhelming probability} that entropy will be higher in the future and was lower in the past, and the sense of probability is   specified by bullet point \#4. There are two ways to understand the measure: 
\begin{itemize}
  \item A measure of probability: the natural measure picks out the correct probability measure of the initial condition. This interpretation yields the Statistical Postulate. 
  \item A measure of typicality: the natural measure is a simple representer of a vague ``collection'' of measures that are equivalent as the measure of typicality of the initial condition. This interpretation yields the Typicality Postulate. 
\end{itemize}
PH together with the Statistical Postulate supports the following classical-mechanical version of the Second Law of Thermodynamics (this is adapted from the Mathematical Second Law described in \citep{goldstein2019gibbs} \S5.2): 

\begin{description}
  \item[The Second Law for $X$] At $t_0$, the actual phase point of the universe $X_0$ starts in a low-entropy macrostate and, with overwhelming probability, it evolves towards macrostates of increasingly higher entropy until it reaches thermal equilibrium, except possibly for entropy decreases that are infrequent, shallow, and short-lived; once $X_t$ reaches $\Gamma_{eq}$, it stays there for an extraordinarily long time, except possible for infrequent, shallow, and short-lived entropy decreases. 
  \end{description}
The Second Law can be stated also in the language of typicality. For simplicity, I will conduct the discussion below mostly in the language of probability. 

The Second Law above is stated for the behavior of the universe, but it also makes plausible  what \cite{goldstein2019gibbs} call a `development conjecture' about isolated subsystems in the universe: 
\begin{description}
  \item[Development Conjecture] Given PH, an isolated system that, at a time $t$ before thermal equilibrium of the universe, has macrostate $\nu$ appears macroscopically in the future, but not the past, of $t$ like a system that at time $t$ is in a typical microstate compatible with $\nu$. 
\end{description}


Classical mechanics with just the fundamental dynamical laws (expressed in equations (\ref{HE})) are time-symmetric. Introducing the probability measure takes care of the Easy Part of the problem of irreversibility, but  to  solve the Hard Part of the problem---the retrodiction to the past, we need to explicitly introduce something that breaks the time symmetry. PH is a simple postulate that does the job.   The  bullet points about  energy shell, macrostate partition, unique correspondence, and the dominance of thermal equilibrium are supposed to follow from the basic postulates about fundamental dynamics (including the structure of the Hamiltonian function) and the probability measure. Adapting the terminology of \cite{albert2015after} and \cite{loewer2016mentaculus}, we call the collection of basic postulates the \textit{Classical Mentaculus}:

\begin{tcolorbox}
\centerline{\textbf{The Classical Mentaculus}}
\begin{enumerate}
\item \textbf{Fundamental Dynamical Laws (FDL)}: the classical microstate of the universe is represented by a point in phase space that obeys the Hamiltonian equations of motion described in equations (\ref{HE}).
 \item \textbf{The Past Hypothesis (PH)}: at a temporal boundary of the universe, the microstate of the universe lies inside $M_0$, a low-entropy macrostate that, given a choice of C-parameters, corresponds to $\Gamma_0$, a small-volume set of points on phase space that are macroscopically similar. 
\item \textbf{The Statistical Postulate (SP)}: given the macrostate $M_0$, we postulate a uniform probability distribution (with respect to the standard Lebesgue measure) over the microstates compatible with $M_0$.
\end{enumerate}
\end{tcolorbox}

This account therefore assigns probability 1 to the initial macrostate. If the probability distribution is given a status of objective probability,  it delivers more than just   the  Second Law. It provides  an exact probability for any proposition formulable in the language of phase space. This is the reason that Albert and Loewer regard the Mentaculus as providing a ``probability map of the world.'' Hence, the Mentaculus has an ambitious scope: it is possible to recover all the non-fundamental regularities, including the special science laws (such as laws of economics), and other arrows of time such as the epistemic arrow, the records arrow, the influence arrow, and the counterfactual arrow. 

Whether the Albert-Loewer project can succeed in their ambitious goal of recovering all the non-fundamental regularities and arrows of time is an interesting question. Nonetheless, the Classical Mentaculus as formulated provides an underpinning for the thermodynamic arrow of time. Given the universality and importance of the Second Law, the Mentaculus should be taken as a serious contender for a promising framework of the structure of a fundamental physical theory. In the next subsection, I examine how to adapt the Classical Mentaculus to the quantum domain. In \S3, I discuss the suggestion that PH should be taken as a candidate fundamental law and some ramifications of the more ambitious project.

\subsection{The Quantum Mentaculus}

Let us  turn to the Boltzmannian account of quantum statistical mechanics from the ``individualist viewpoint.'' Consider  a quantum-mechanical system with $N$ fermions (with $N> 10^{20}$) in a box $\Lambda = [0, L]^3 \subset \mathbb{R}^3$ and a Hamiltonian $\hat{H}$. (Here, I follow the discussions in \citep{goldstein2010approach} and \citep{goldstein2010approachB}.)

\begin{enumerate}
\item Microstate: at any time $t$, the microstate of the system is given by a normalized (and anti-symmetrized) wave function:
\begin{equation}
\psi(\boldsymbol{q_1}, ..., \boldsymbol{q_N}) \in \mathscr{H}_{total} = L^2 (\Lambda^{N}, \mathbb{C}^k) \text{ , } \parallel \psi \parallel_{L^2} = 1,
\end{equation}
where $\mathscr{H}_{total} = L^2 (\Lambda^{N},  \mathbb{C}^k)$ is the total Hilbert space of the system. 

\item Dynamics: the time dependence of $\psi(\boldsymbol{q_1}, ..., \boldsymbol{q_N}; t)$ is given by the Schr\"odinger equation: 
\begin{equation}\label{SE}
 i\hbar \frac{\partial \psi}{\partial t} = \hat{H} \psi.
\end{equation}

\item Energy shell: the physically relevant part of the total Hilbert space is the subspace (``the energy shell''):
\begin{equation}
\mathscr{H} \subseteq \mathscr{H}_{total} \text{ , } \mathscr{H} = \text{span} \{ \phi_\alpha : E_\alpha \in [E, E+\delta E ]  \},
\end{equation}
This is the subspace (of the total Hilbert space) spanned by energy eigenstates $\phi_\alpha$ whose eigenvalues $E_\alpha$ belong to the $[E, E+\delta E]$ range.  Let $D = \text{dim} \mathscr{H}$, the number of energy levels between $E$ and $E+\delta E$. 

We only consider wave functions $\psi$ in $\mathscr{H}$.

\item Measure: given a subspace $\mathscr{H}$, the measure $\mu_S$ is  the surface area measure on the unit sphere in that subspace $\mathscr{S}(\mathscr{H})$.\footnote{For simplicity, let us assume that the subspaces we deal with are finite-dimensional. In cases where the Hilbert space is infinite-dimensional, it is an open and challenging technical question.  For example, we could use Gaussian measures in infinite-dimensional spaces, but we  no longer  have uniform probability distributions. }

\item Macrostate: with a choice of macro-variables,\footnote{For technical reasons,  \cite{von1955mathematical}  suggests that we round up these macro-variables (represented by quantum observables) so as to make the  observables commute. See \citep[section 2.2]{goldstein2010long} for a discussion of von Neumann's ideas. } the energy shell $\mathscr{H}$ can be orthogonally decomposed into macro-spaces (subspaces):
\begin{equation}
\mathscr{H} = \oplus_\nu \mathscr{H}_\nu \text{ , } \sum_\nu \text{dim}\mathscr{H}_\nu  = D
\end{equation}
Each $\mathscr{H}_\nu$ corresponds  to small ranges of values of macro-variables that are chosen in advance. 

Caveat: similarly to the classical case, the decomposition of Hilbert space into macrostates requires some stipulation of the exact values of the C-parameters. But in the quantum case, these parameters includes coarse-graining sizes, correspondences of functions, and also the cut-off values of how much support a quantum state needs to be inside a subspace to be counted towards belonging to the macrostate (see the next bullet point). Without the exact choices of the C-parameters, the decomposition is inexact and it is vague which microstate belongs to which macrostate. Again, it is also expected that given the nature of the actual forces, some decompositions will be superior to others for supporting generalizations in the special sciences.

\item Non-unique correspondence: typically, a wave function is in a superposition of macrostates and is not entirely in any one of the macrostates (even if we represent macrostates with  exact subspaces). However, we can make sense of situations where $\psi$ is (in the Hilbert space norm) very close to a macrostate $\mathscr{H}_\nu$: 
\begin{equation}\label{close}
\bra{\psi} P_{\nu}  \ket{\psi} \approx 1,
\end{equation}
where $P_{\nu}$ is the projection operator onto $\mathscr{H}_{\nu}$. This means that  $\ket{\psi}$ lies almost entirely in $\mathscr{H}_{\nu}$. In this case, we say that $\ket{\psi}$ is in macrostate $\nu$. 

\item Thermal equilibrium: typically, there is a dominant macrostate $\mathscr{H}_{eq}$ that has a dimension that  is almost equal to D: 
\begin{equation}
\frac{\text{dim} \mathscr{H}_{eq}}{\text{dim} \mathscr{H}} \approx 1.
\end{equation}
A system with wave function $\psi$ is in equilibrium if the wave function $\psi $ is very close to $\mathscr{H}_{eq}$ in the sense of (\ref{close}):  $\bra{\psi} P_{eq}  \ket{\psi} \approx 1.$


\item Boltzmann Entropy: the Boltzmann entropy of a quantum-mechanical system with wave function $\psi$ that is in macrostate $\nu$ is given by:
\begin{equation}\label{Boltzmann}
S_B (\psi) = k_B \text{log} (\text{dim} \mathscr{H}_\nu ),
\end{equation}
where $\mathscr{H}_\nu$ denotes the subspace containing almost all of $\psi$ in the sense of (\ref{close}). The thermal equilibrium state thus has the maximum entropy: 
\begin{equation}
S_B (eq) = k_B \text{log} (\text{dim} \mathscr{H}_{eq} ) \approx  k_B \text{log} (D),
\end{equation}
where \emph{eq} denotes the equilibrium macrostate. 

\item Low-Entropy Initial Condition: on the assumption that we can model the universe as a quantum-mechanical system, let us postulate a special low-entropy boundary condition on the universal wave function---the quantum-mechanical version of  PH: 
\begin{equation}\label{QuantumPH}
\Psi(t_0) \in \mathscr{H}_{PH} \text{ , } \text{dim} \mathscr{H}_{PH} \ll \text{dim}\mathscr{H}_{eq} \approx \text{dim} \mathscr{H}
\end{equation}
where $\mathscr{H}_{PH}$ is the PH macro-space with dimension much smaller than that of the equilibrium macro-space.\footnote{Again, we  assume that $\mathscr{H}_{PH}$ is finite-dimensional, in which case we can use the surface area measure on the unit sphere as the typicality measure for \# 10. It remains an open question in QSM about how to formulate the low-entropy initial condition when the initial macro-space is infinite-dimensional.} Hence, the initial state has very low entropy in the sense of (\ref{Boltzmann}). More details can be added to narrow down the range of choices of  $\mathscr{H}_{PH}$. 

\end{enumerate}

The quantum Boltzmannian account is similar to the classical one. The higher-entropy macrostates have much higher dimensions than lower-entropy ones, and the equilibrium macrostate has by far the largest dimension. It is plausible that, unless the dynamics is very contrived, a medium-entropy wave function will find its way through larger and larger subspaces and eventually arrive at the  equilibrium subspace. Again, if typical wave functions in non-equilibrium macrostates evolve towards higher entropy in the future, then typical ones also come from higher entropy states in the past. The quantum mechanical Past Hypothesis blocks that inference. The reason there is a thermodynamic arrow in a quantum universe is because the universal wave function started in a special state, a subspace with very low entropy (the one described by the quantum PH). Assuming the quantum PH, it is reasonable to expect that \textit{with overwhelming probability} the entropy is higher in the future and lower in the past, with the probability measure specified in bullet point \#4. Again, we can understand it as a measure of probability or a measure of typicality. 

Together, the probability distribution and PH support the quantum-mechanical version of the Second Law: 

\begin{description}
  \item[The Second Law for $\Psi$] At $t_0$, the actual wave function of the universe $\Psi_0$ starts in a low-entropy macrostate, and with overwhelming probability, it evolves towards macrostates of increasingly higher entropy until it reaches thermal equilibrium, except possibly for entropy decreases that are infrequent, shallow, and short-lived; once $\Psi_t$ reaches $\mathscr{H}_{eq}$, it stays there for an extraordinarily long time, except possibly for infrequent, shallow, and short-lived entropy decreases. 
  \end{description}
(This makes plausible a similar Development Conjecture for typical isolated subsystems.) 

Note again we stipulate three basic postulates in the quantum version of the Boltzmannian account: the fundamental dynamical laws, PH, and SP. Let us call this the Quantum Mentaculus: 

\begin{tcolorbox}
\centerline{\textbf{The Quantum Mentaculus}}
\begin{enumerate}
\item \textbf{Fundamental Dynamical Laws (FDL):} the quantum microstate of the universe is represented by a wave function $\Psi$ that obeys the Schr\"odinger equation (\ref{SE}). 
\item \textbf{The Past Hypothesis (PH)}: at a temporal boundary of the universe, the wave function $\Psi_0$ of the universe lies inside a low-entropy macrostate that, given a choice of C-parameters,  corresponds to $\mathscr{H}_{PH}$, a low-dimensional subspace of the total Hilbert space. 
\item \textbf{The Statistical Postulate (SP)}: given the subspace $\mathscr{H}_{PH}$, we postulate a uniform probability distribution (with respect to the surface area measure on the unit sphere of $\mathscr{H}_{PH}$) over the wave functions compatible with $\mathscr{H}_{PH}$.
\end{enumerate}
\end{tcolorbox}

The Quantum Mentaculus, as a candidate fundamental theory of physics, faces the quantum measurement problem. To solve the measurement problem, there are three promising options: Everettian quantum mechanics, Bohmian mechanics, and GRW spontaneous collapse theories. We have three distinct kinds of the Quantum Mentaculus. 

First, the Everettian version is completely the same as the original Quantum Mentaculus in terms of the basic postulates. However, it diverges greatly from common sense: we have to give up the  expectation that experimental outcomes are unique and determinate. Instead, our experiences are to be understood as experiences of agents in an emergent multiverse  \citep{wallace2012emergent}. 
 
Second, the Bohmian version posits that in addition to the wave function, which evolves unitarily according to the Schr\"odinger equation, particles have precise locations, and their configuration $Q = (Q_1, Q_2, ... , Q_N)$ follows the guidance equation, which is an additional law in the theory:  

\begin{equation}\label{GE}
 \frac{dQ_i}{dt} = \frac{\hbar}{m_i} \text{Im} \frac{ \nabla_i \psi (q) }{  \psi (q)} (q=Q)
\end{equation}
Moreover, the initial particle distribution is given by the quantum equilibrium distribution: 
\begin{equation}\label{QEH}
\rho_{t_0} (q) = |\psi(q, t_0)|^2
\end{equation}
Adding the above  two  postulates to the Quantum Mentaculus completes the Bohmian Mentaculus. 

Third, the GRW version requires revisions to the linear evolution represented by the Schr\"odinger equation. The wave function typically obeys the Schr\"odinger equation, but the linear evolution is interrupted randomly (with rate $N\lambda$, where $N$ is the number of particles and $\lambda$ is a new constant of nature of order $10^{-15}$ s$^{-1}$) by collapses: 
\begin{equation}\label{WFcollapse}
\Psi_{T^+} = \frac{\Lambda_{k} (X)^{1/2} \Psi_{T^-} }{||  \Lambda_{k} (X)^{1/2} \Psi_{T^-}  || },
\end{equation}
where $\Psi_{T^-} $ is the pre-collapse wave function,  $\Psi_{T^+} $ is the post-collapse wave function, the collapse center $X$ is chosen randomly with probability distribution $\rho(x) = ||  \Lambda_{k} (x)^{1/2} \Psi_{T^-}  ||^2 dx$,  $k \in \{1, 2, ... N\}$ is chosen  randomly with uniform distribution on that set of particle labels, and the collapse rate operator is defined as:
\begin{equation}\label{collapserate}
\Lambda_{k} (x) = \frac{1}{(2\pi \sigma^2)^{3/2}} e^{-\frac{(Q_k -x)^2}{2\sigma^2}},
\end{equation}
where $Q_k$ is the position operator of ``particle'' $k$, and $\sigma$ is another new constant of nature of order $10^{-7}$ m postulated in current GRW theories. The GRW Mentaculus replaces the deterministic Schr\"odinger evolution of the wave function by this stochastic process. It still requires PH. However, as Albert (2000) \S7 points out, it is plausible (though not proven) that SP is no longer needed, and the GRW collapses suffice to make anti-entropic trajectories unlikely (through the quantum probabilities stipulated by the GRW stochastic process). (See  Ismael's contribution in this volume.)

\subsection{The Wentaculus}

In this subsection, let us consider the Boltzmannian account of quantum statistical mechanics with a very special ``fundamental density matrix.'' This account is inspired by \citep{durr2005role}, proposed in \citep{chen2018IPH}, and discussed at length in \citep{chen2019quantum1, chen2018valia}. It is another variation on the same theme from Boltzmann, but it suggests some astonishing possibilities,  one of which is the Initial Projection Hypothesis that will be introduced shortly. 

The density matrix can play the same \textit{dynamical role} as the wave function does in the previous theories. In a quantum system represented by a density matrix $W$, $W$ is the complete characterization of the quantum state; it does not necessarily refer to a statistical state representing our ignorance of the underlying wave function. In general, $W$ can be a pure state or a mixed state. A density matrix $\hat{W}$ is pure if $\hat{W} = \ket{\psi} \bra{\psi}$ for some $\ket{\psi}$. Otherwise it is mixed.  For a spinless $N$-particle quantum system,  a density matrix of the system is a positive, bounded, self-adjoint operator $\hat{W} : \mathscr{H} \rightarrow \mathscr{H}$ with $\text{tr} {\hat{W}} = 1$, where $\mathscr{H}$ is the Hilbert space of the system. In terms of the configuration space $\mathbb{R}^{3N}$, the density matrix can be viewed as a function $W: \mathbb{R}^{3N} \times \mathbb{R}^{3N} \rightarrow \mathbb{C}$. In the unitary case, $\hat{W}$ always evolves deterministically according to the von Neumann equation: 
\begin{equation}\label{VNM}
i \hbar \frac{d \hat{W}(t)}{d t} = [\hat{H},  \hat{W}].
\end{equation}
Equivalently: 
\begin{equation}\label{VNMQ}
i \hbar \frac{\partial W(q, q', t)}{\partial t} = \hat{H}_q W(q, q', t) -  \hat{H}_{q'} W(q, q', t),
\end{equation}
where  $\hat{H}_{q}$ means that the Hamiltonian  $\hat{H}$ acts on the variable $q$. The von Neumann equation generalizes  the Schr\"odinger equation (\ref{SE}). 

Importantly, now the ``fundamental'' quantum state can be either pure or mixed. Even when it is mixed, there is no underlying pure state that is more basic. The mixed state is completely objective. This perspective, called \emph{Density Matrix Realism}, is in sharp contrast to the prevalent view called \emph{Wave Function Realism}. 
In  the density-matrix realist framework, we need to modify the Boltzmannian quantum statistical mechanics described in the earlier section. Here are the key changes: 
\begin{itemize}
\item Microstate: at any time $t$, the microstate of the system is given by a density matrix $\hat{W}(t)$ that can be pure or mixed. (Macrostates are still represented by orthogonal subspaces of the energy shell.)
\item Dynamics: in the unitary case, the density matrix $\hat{W}(t)$ evolves according to the von Neumann equation (\ref{VNM}). 
\item Being in a macrostate: typically, a density matrix is  a superposition of macrostates and is not entirely in any one of the macrospaces. However, we can make sense of situations where $\hat{W}(t)$ is very close to a macrostate $\mathscr{H}_\nu$: 
\begin{equation}\label{Wclose}
\text{tr} (\hat{W}(t) I_{\nu})  \approx 1,
\end{equation}
where $I_{\nu}$ is the projection operator onto $\mathscr{H}_{\nu}$. This means that almost all of $\hat{W}(t)$ is in $\mathscr{H}_{\nu}$. In this situation, we say that $\hat{W}(t)$ is in macrostate $\mathscr{H}_{\nu}$. 
\item Thermal equilibrium: typically, there is a dominant macro-space $\mathscr{H}_{eq}$ that has a dimension that  is almost equal to D: 
\begin{equation}
\frac{\text{dim} \mathscr{H}_{eq}}{\text{dim} \mathscr{H}} \approx 1.
\end{equation}
A system with density matrix $\hat{W}(t)$ is in equilibrium if  $\hat{W}(t)$ is very close to $\mathscr{H}_{eq}$ in the sense of (\ref{Wclose}):  $\text{tr} (\hat{W}(t) I_{eq})  \approx 1$.
\item Boltzmann entropy: the Boltzmann entropy of a quantum-mechanical system with density matrix $\hat{W}(t)$ that is very close to a macrostate $\nu$ is given by:
\begin{equation}\label{Boltzmann}
S_B (\hat{W}(t)) = k_B \text{log} (\text{dim} \mathscr{H}_\nu ),
\end{equation}
for which $W$ is in  macrostate $\mathscr{H}_\nu$  in the sense of (\ref{Wclose}). 
\end{itemize}
Next, let us consider how to adapt PH in a density-matrix realist framework. The wave-function version of PH says that every initial wave function is entirely contained in the PH subspace $\mathscr{H}_{PH}$. Similarly, for density-matrix theories, we can  propose that every initial density matrix is entirely contained in the PH subspace: 
\begin{equation}\label{WPH}
\text{tr} (\hat{W}(t_0) I_{PH}) =1  \text{ , } \text{dim} \mathscr{H}_{PH} \ll \text{dim}\mathscr{H}_{eq} \approx \text{dim} \mathscr{H}
\end{equation}
where $I_{PH}$ is the projection operator onto the PH subspace. Assuming $\mathscr{H}_{PH}$ is finite-dimensional, there is also a natural probability distribution over all density matrices inside this subspace. See \citep{chen2020uniform} for a mathematical characterization. The probability distribution and the density-matrix Past Hypothesis support a Second Law for $W$, which is similar to the Second Law for $\Psi$. 

However, there is an even more natural way to implement the idea of PH in the density-matrix framework, which I favor. PH picks out a particular subspace $\mathscr{H}_{PH}$. It is canonically associated with its projection $I_{PH}$. In matrix form, it can be represented as a block-diagonal matrix that has a $k\times k$ identity block, with $k = \text{dim}{\mathscr{H}_{PH}}$, and zero everywhere else. There is a natural density matrix associated with $I_{PH}$, namely the normalized projection $\frac{I_{PH}}{\text{dim} \mathscr{H}_{PH}}$. Hence, we have picked out the natural density matrix associated with the PH subspace. I propose that the initial density matrix is the normalized projection onto $\mathscr{H}_{PH}$:
\begin{equation}\label{IPH}
\hat{W}_{IPH}(t_0) = \frac{I_{PH}}{\text{dim} \mathscr{H}_{PH}}.
\end{equation}
I  call this postulate the \emph{Initial Projection Hypothesis} (IPH) in \citep{chen2018IPH}.  Crucially, it is different from (\ref{QuantumPH}) and (\ref{WPH});  while IPH picks out a unique  quantum state given PH, the other two permit infinitely many possible  quantum states inside the PH subspace. Remarkably, we no longer need a fundamental postulate about probability or typicality for the quantum state. We know that we can decompose a density matrix \textit{non-uniquely} into a probability-weighted average of pure states, and in the canonical way we can decompose $\hat{W}_{IPH}(t_0)$ as an integral of pure states on the unit sphere of $\mathscr{H}_{PH}$ with respect to the  uniform probability distribution:
\begin{equation}\label{MacroW}
\hat{W}_{IPH} (t_0) = \int_{\mathscr{S}(\mathscr{H}_{PH})} \mu(d\psi) \ket{\psi} \bra{\psi}.
\end{equation}
The decomposition here is not an intrinsic expression of what $\hat{W}_{IPH} (t_0)$ is, as witnessed by the non-uniqueness. But the expression is something that can nonetheless be used fruitfully in statistical analysis \citep[section 3.2.3]{chen2018valia}. 

By doing away with the need for an extra postulate about initial quantum states, we only  need two basic postulates.  I call the theory \textit{the Wentaculus}: 

\begin{tcolorbox}
\centerline{\textbf{The Wentaculus}}
\begin{enumerate}
\item \textbf{Fundamental Dynamical Laws (FDL):} the quantum state of the universe is represented by a density matrix  $\hat{W}(t)$ that obeys the von Neumann equation (\ref{VNM}).
\item \textbf{The Initial Projection Hypothesis (IPH)}: at a temporal boundary of the universe, the density matrix is the normalized projection onto $\mathscr{H}_{PH}$,  a low-dimensional subspace of the total Hilbert space. (That is, the initial quantum state of the universe is $\hat{W}_{IPH} (t_0)$ as described in equation (\ref{IPH}).)
\end{enumerate}
\end{tcolorbox}

Similar to the Quantum Mentaculus, the Wentaculus also suffers from the quantum measurement problem. There are three promising solutions, each of which gives rise to a distinct version of the Wentaculus. 

First, there is the Everettian Wentaculus that looks exactly like the basic Wentaculus. For this theory, we need to embrace the idea that there is a (vague) multiplicity of emergent worlds that is similar to the original Everettian theory. What is interesting about the Everettian Wentaculus is that it suggests an astonishing possibility. If IPH is interpreted as a fundamental law, then the theory is \textit{strongly deterministic}, in the sense of \citep{roger1989emperor} that the laws pick out a unique micro-history of the fundamental ontology (represented by $W(t)$). This theory does not postulate any objective probability. This is another step towards the Everettian aspiration of constructing a theory without any fundamental contingency. 

Second, the Bohmian Wentaculus postulates that, in addition to the universal density matrix $W$ that evolves unitarily according to the von Neumann equation, there are actual particles that have precise locations in physical space, represented by $\mathbb{R}^3$. The particle configuration $Q = (Q_1, Q_2, ... , Q_N) \in \mathbb{R}^{3N}$ follows the guidance equation (written for the $i$-th particle):\footnote{This version of the guidance equation is first proposed by \cite{bell1980broglie}, then recast as the dynamical equation for the fundamental density matrix in \cite{durr2005role}.} 
\begin{equation}\label{WGE}
\frac{dQ_i}{dt} = \frac{\hbar}{m_i} \text{Im} \frac{\nabla_{q_{i}}  W (q, q', t)}{ W (q, q', t)} (q=q'=Q),
\end{equation}
 Moreover, the initial particle distribution is given by the density-matrix version of the quantum equilibrium distribution: 
\begin{equation} \label{WQEH}
P(Q(t_0) \in dq) = W (q, q, t_0) dq.
\end{equation}

Third, the GRW Wentaculus postulates that the universal density matrix typically obeys the von Neumann equation, but the linear evolution is interrupted randomly by collapses (with rate $N\lambda$, where $N$ is the number of particles and $\lambda$ is a new constant of nature of order $10^{-15}$ s$^{-1}$):\footnote{To my knowledge, the W-GRW equations first appear in \citep{allori2013predictions}. }
\begin{equation}\label{collapse}
W_{T^+} = \frac{\Lambda_{k} (X)^{1/2} W_{T^-} \Lambda_{k} (X)^{1/2}}{\text{tr} (W_{T^-} \Lambda_{k} (X)) },
\end{equation}
where $W_{T^-} $ is the pre-collapse density matrix, $W_{T^+} $ is the post-collapse density matrix, with $k$ uniformly distributed in the $N$-element set of particle labels and $X$ distributed by  $\rho(x) = \text{tr} (W_{T^-} \Lambda_{k} (x)),$
with the collapse rate operator defined as before in (\ref{collapserate}).

In this section, we presented several versions of PH. They can all be traced back to the original Boltzmannian idea that the initial state of the universe is special and has low entropy.  Differences arise when we move to the Wentaculus framework where IPH selects a unique and simple initial \textit{microstate} of the universe. The microstate is given by a mixed-state density matrix. The reason a mixed state can play the role of a microstate is because it enters directly into the fundamental dynamical equations, such as  (\ref{VNM}), (\ref{WGE}), and (\ref{collapse}).  The different initial-condition postulates---(\ref{PH}), (\ref{QuantumPH}), and (\ref{WPH})---form a family, and I shall continue using the generic label, the ``Past Hypothesis,'' to refer to them and will only use specific labels when their differences are relevant.

\section{Why the Past Hypothesis is Law-Like}

It is clear that PH has a special status in the Boltzmannian account. It has been suggested that PH is like a law of nature. For example, this is emphasized by \cite{feynman2017character}[1965] as quoted in the epigraph. 
Making a similar point about classical statistical mechanics, \cite{goldstein2019gibbs} suggests that PH is an interesting kind of law:  
\begin{quotation}
	The past hypothesis is the one crucial assumption we make in addition to the dynamical laws of classical mechanics. The past hypothesis may well have the status of a law of physics---not a dynamical law but a law selecting a set of admissible histories among the solutions of the dynamical laws.
\end{quotation}

In this section, I offer four types of positive arguments to support the view that PH is a candidate fundamental law of nature. These arguments also support the weaker thesis that PH is a candidate axiom in the fundamental theory. 
My methodology is naturalistic and functionalist. I argue for the Nomic Status of PH by locating the roles of the fundamental laws in our physical theories  and by showing that PH plays such roles.   These roles include  backing  scientific explanations, constraining nomological possibilities, and supporting objective probabilities. 

Some of these arguments (\S3.1--3.3) are related to ideas that have appeared in the literature. I try to make the premises explicit, with the hope that the arguments are clear enough for others who disagree to examine and criticize. Usually, the  arguments are made in the Humean framework, but as I argue, they can also be made on behalf of non-Humeans who have a minimalist conception of what it is for laws to really \textit{govern}. That is the account I favor.
Of course, the minimalist account is at odds with the idea about ``dynamical governing'':
\begin{description}
  \item[Dynamical Governing] Only dynamical laws can be fundamental laws of nature.
\end{description}
In \S3.4, I offer a new argument for the Nomic Status of PH based on considerations of the nature of quantum entanglement.

\subsection{Arguments from the Second Law}

The (fundamental) nomic status of PH is supported by the nomic status of the Second Law of Thermodynamics.   The Second Law is a law of nature; whatever underlies a law is a law.  A law that cannot be derived from other laws is a fundamental law; therefore, PH is a fundamental law. Let us spell out the argument in more detail:  

\begin{description}
  \item[P1] The Second Law of Thermodynamics is a law of nature. 
  \item[P2] A law of nature can be scientifically explained only by appealing to more fundamental laws of nature and laws of mathematics. 
  \item[P3] The Second Law of Thermodynamics is scientifically explained (in part) by PH, and PH is not a law of mathematics.
  \item[C1] So, PH is a law of nature and is more fundamental than the Second Law.
  \item[P4] PH is not scientifically explained by  fundamental laws. 
  \item[P5] A law of nature that is not scientifically explained by fundamental laws is a fundamental law.
  \item[C2] So, PH is a fundamental law of nature. 
\end{description}

Comments on P1. First, the Second Law of Thermodynamics summarizes an important  regularity:  the tendency for things to become more chaotic and more decayed as time passes. It is part of our concept of lawhood that this irreversible tendency is law-like. We learn about this law much more directly in our experiences than the microscopic equations of motion. Second, nature's irreversible tendency is  encoded in our concept of physical necessity. For example, we learn that it is physically impossible for a metal rod to spontaneously heat up on one side and then cool down on the other;  it is physically impossible to create a perpetual motion machine of the second kind (a machine that violates the Second Law), and this is impossible no matter who tries to do it -- whenever and wherever.\footnote{This becomes more complicated  if an ``Albertian demon'' turns out to be physically possible. See (Albert, 2000, \S5) and Maudlin's  contribution in this volume.} Hence, the Second Law is not an accidental feature of the world.  Of course, the usual formulation of the Second Law in terms of the absolute monotonic increase of entropy is too strong. It should be modified in two ways: it holds for the overwhelming majority of nomologically possible initial conditions, and for each entropic trajectory, there can be short-lived, shallow, and infrequent decreases of entropy (see Second Laws for $X$ and for $\Psi$). 

Recognizing the importance of the Second Law, \cite{EddingtonNPW} suggests:
\begin{quotation} If someone points out to you that your pet theory of the universe is in disagreement with Maxwell’s equations---then so much the worse for Maxwell's equations. If it is found to be contradicted by observation---well, these experimentalists do bungle things sometimes. But if your theory is found to be against the second law of thermodynamics I can give you no hope; there is nothing for it but to collapse in deepest humiliation. \end{quotation}
That may be too strong. Nevertheless, we should accept that the Second Law is nomologically necessary. It is not merely an accidental feature of the world, such as the contingent fact that all gold spheres are less than one mile in diameter. Third, counterfactuals are backed by laws of nature; laws are what we hold fixed when evaluating counterfactuals. The Second Law  backs counterfactuals about macroscopic processes that display a temporal asymmetry: if there were a half-mixed ink drop in my water cup right now, it would have been more separated in the past and more evenly mixed in the future. (See \S3.2 for more on the counterfactual arrow.) 

Comments on P2. The notion of scientific explanation here is not a fully analyzed notion. What is relevant to this argument is that the Second Law is supposed to be \textit{derived as a theorem} from the basic postulates of the Mentaculus or the Wentaculus. It is expected that, assuming the  laws of mathematics, the fundamental dynamical laws, PH, and SP,  an initial microstate starting from the initial macrostate will, with overwhelming probability, travel to macrostates of increasingly higher entropy until it reaches thermal equilibrium (except possibly for short-lived and infrequent decreases of entropy). So, it is the Second Law's mathematical derivation from the basic postulates of the physical theory that is the relevant notion here.\footnote{This is  different from the notion of metaphysical explanation. See \citep{loewer2012two}, \citep{hicks2015humean}. On their views, a fundamental law of nature is metaphysically explained (but not scientifically explained) by the matter distribution.} 

A standard response to P2 is that a non-fundamental law (such as those in the special sciences) can be explained (in part) by some contingent boundary conditions.  Examples may include the laws of genetics and laws of economics. However, it is not clear what does the explanatory work in those cases. Let us suppose that  some special science law $S$ arise from boundary conditions $B$. Suppose $B$ obtains. Then there is a high objective probability that $S$ obtains. What is the origin of these objective probabilities?  What is the physical explanation? If everything is ultimately physical, and the physical theory is informationally complete in so far as the motion of objects go, then it seems that the objective probabilities are ultimately backed by some postulates in physics. The probabilities in physics may supply conditional probabilities on which $Pr(S|B)$ is high. What does the explaining, then, is the probability supplied by physics, and the real law should be the high probability of $S$ obtaining given $B$, which is consistent with the physicalist picture we started with.  

Winsberg (this volume) offers another potential counterexample to P2. He suggests that due to the near certainty of the existence of Boltzmann brains and large fluctuations in future epochs of the universe, it is important to postulate also a Near Past Hypothesis (see also \citep{chen2020time}):

\begin{description}
  \item[Near Past Hypothesis (NPH)] We are inside the first epoch of the universe between the initial time and the first relaxation to thermal equilibrium. 
\end{description}
Winsberg argues that NPH should have the same status as  PH and SP because it is also a necessary postulate to derive the Second Law (among other special science laws). However, NPH is an indexical statement about our location in time; as such, it cannot be a candidate fundamental law of nature. So if a non-law can be part of the explanation for the Second Law, then there is no reason to think that only laws can back laws. This is an interesting insight, but I do not think Winsberg's argument undermines P2. If indexical statements cannot be laws, then the Second Law should not be stated in an indexical way. We should use the non-indexical version of the Second Law (and special science laws), such as the Second Law for $X$, the Second Law for $\Psi$, and the Second Law for $W$. In those versions, fluctuations are already taken into account (in a non-indexical way). Hence, we do not need to invoke NPH to derive those versions of the Second Law from PH, SP, and the dynamical laws.

Comments on P3. This premise is true if we grant the explanatory success of the Boltzmannian account, which is assumed in this paper. It is clear that PH is not a law of mathematics. 

Comments on P4.  P4  is an open scientific question. Perhaps some future theory (e.g. along the lines of \citep{carroll2004spontaneous}) can explain PH using some simple and satisfactory dynamical laws. Still, it is also a scientific possibility that PH remains a fundamental law in the final theory and is not explained further. Given the openness of P4, we should accept C2 only to the degree of acknowledging PH as a \textit{candidate} fundamental law. 

Comments on P5. This follows from our concept of a fundamental law of nature.

The argument above supports the (fundamental) nomic status of PH. If Nomic Status implies Axiomatic Status, then the argument also supports the idea that PH is an axiom in the fundamental physical theory. But there is another, more straightforward  argument for the Axiomatic Status.   The predictive consequences of a physical theory should come entirely from its axioms and their deductive consequences. A good physical theory  aims at capturing as many regularities as possible using simple axioms.  The Second Law describes an important regularity. Therefore, we  postulate  PH and SP in addition to the fundamental dynamical laws. These postulates have an axiomatic status in the Mentaculus. 

The Mentaculus is a good theory; it is better than ``Mentaculus-,'' the Mentaculus minus PH and SP. The Mentaculus predicts not only the motion of planets, but also the overwhelming probability that my table will not spontaneously rearrange itself into the shape of a statue. The Mentaculus- can tell us everything about the motion of planets but is silent about many macroscopic regularities we see around us. Even so, the Mentaculus is a pretty simple theory. Someone might suggest that to achieve maximal predictive power, we can add a statement about the exact microstate of the universe at $t_0$ as an additional axiom to Mentaculus-. But the exact microstate is a detailed fact that complicates the Mentaculus - such that its axioms will no longer be simple enough. (In the Wentaculus, IPH pins down a quantum microstate, but its informational content and simplicity level are the same as those of PH in the Mentaculus.)

\subsection{Arguments from Other Asymmetries}

The thermodynamic arrow of time described by the Second Law is best explained by PH. That provides strong support for the Nomic Status and the Axiomatic Status of PH. What about other arrows of time? In this subsection, I present arguments based on the counterfactual arrow, the records arrow, the epistemic arrow, and the intervention arrow. The upshot is that they can also be traced back to the nomic status of PH, without which they would be left completely mysterious. Many of these ideas can be found in (Albert 2000, 2012) and \citep{LoewerCatSLaw}, and they are also discussed in \citep{frisch2005counterfactuals, frischcausation}, \citep{demarest2019mentaculus}, (Fernandes, this volume), \citep{callender2004measures}, \citep{horwich1987asymmetries}, and \citep{reichenbach1956direction}.

\textit{The records arrow.} We have photographs and videos of WWII but no photographs of the next major world war. We have detailed accounts of the life of President Washington but no detailed accounts of the life of the 65th president of the United States. There are craters on the moon indicating past meteorite impacts but no craters indicating future meteorite impacts. Similarly, there are fossils, rocks, ice sheets, all of which tell us the state of our planet in the past, but we do not have similarly abundant records that tell us the state of our planet in the future. 

What is it about our world such that there are abundant records of the past but few, if any, records about the future? One could appeal to some A-theory of time, according to which the future does not yet exist and the past has already happened. So there cannot be records about the future because there are no facts about the future. But this does not seem to provide a satisfactory scientific explanation, as the temporally asymmetric probabilistic correlations have to be accepted as brute facts. In any case, in a block-universe picture compatible with a B-theory of time, the past, present, and future are all equally real; all events exist tenselessly. There are strong probabilistic correlations between physical records (e.g. fossils) that exist at a particular time and physical systems (e.g. dinosaurs) that exist at an earlier time, but no strong correlations between physical records and  events at a later time.\footnote{The notions of ``earlier'' and ``later'' here can be fleshed out in a way that does not refer to an intrinsic arrow of time. What matters here is not there are facts about earlier times or later times but simply that the probabilistic correlations are temporally asymmetric. }

PH offers an explanation.  Albert (2000) suggests that a record is a relation between two temporal ends of a physical process. A record enables us to infer what happens inside the temporal interval. For example, in a lab, the record of an electron passing through a small slit is the relation between the "ready state" of the measuring instrument at $t_1$ and the "click" state of the measuring instrument at $t_2$. If the instrument moves from "ready" to "click," then we can infer that an electron has passed through the slit between $t_1$ and $t_2$. But if the instrument was not at "ready" at $t_1$, we cannot infer that.    However, to know that the instrument was indeed "ready," we also need to rely on an earlier record. This seems to go back in time \textit{ad infinitum}, to records about the lab, and to records about the larger environment, and eventually to records about earlier states of the universe. To know that the cosmic microwave background (CMB) data is reliable, we also need to postulate that there is some "ready" state at the beginning of the universe. PH, stipulated at (or around) $t_0$, is the "mother of all ready states." It provides the underpinning necessary for our inferences based on records to be carried out.\footnote{More details are needed to fully explain the records asymmetry;  \cite{rovelli2020memory} provides an interesting analysis that adds additional constraints on the initial condition. } 

However, it is not enough that PH be true. We also need to justify the important fact that physical records are reliable. For this, we need  PH to have the status of a law.  (If PH is not derived from other laws, it will have the status of a fundamental law.) If a theory predicts that  it is unlikely for physical systems that look like records to reliably indicate past events, then the theory would undermine the rational justification for believing in it. Such a theory would be \textit{epistemically self-undermining} because we believe in physical theories based on records about past experiments and observations.\footnote{This is somewhat parallel to the situation of empirical adequacy and records in Everettian quantum mechanics. See \citep{barrett1996empirical} on the latter.} The Mentaculus without PH is such a theory. It would predict that most "records" come about from random fluctuations. If we dig out a shoe of Napoleon, most likely it came about from random fluctuations and not from a low-entropy, past state. Postulating PH as a law avoids that. The uniform probability distribution conditionalized on PH will predict that most physical systems that look like "records" will be reliable records about the past (here we set aside the problems of large future fluctuations).

\textit{The epistemic arrow.} Given some information about the present, there is some sense in which our knowledge about the past is more vast,  detailed, and easily gained than our knowledge about the future. We know that the sun will (likely) rise tomorrow, but we do not know who will win the US presidential election of 2028 and when exactly the stock market will crash over the next 20 years. But we know exactly who won the election of 1860, when exactly the stock market crashed in the last 20 years, and so on.   Similar to the records arrow, the epistemic arrow is especially puzzling in a block-universe picture compatible with a B-theory of time. All facts about the past, present, and future are out there. Why is our knowledge so skewed towards one temporal direction? 

Albert (2000) explains the epistemic arrow in terms of the records arrow, which in turn is explained by the Nomic Status of PH.  The basic idea is this. We distinguish between inferences based on records from inferences based on predictions or retrodictions. The latter uses only the current macrostate together with the dynamical laws plus SP (construed as an unconditionalized uniform probability distribution on the energy shell of phase space) to the past (retrodictions) and to the future (predictions). Inferences based on predictions will tell us that with overwhelming probability, the sun will rise tomorrow and the ice cubes in my coffee will melt in the next hour, but inferences based on retrodictions will get most things  wrong about the past. For example, retrodictions will tell us that the ice cubes in my coffee were actually smaller in the past (they spontaneously got larger in my coffee), and all  the books about someone named Lincoln winning the 1860 election came about from random collisions of particles. However, inferences based on records are much more powerful and demand much less detailed information about the current macrostate of the world. We can infer to past states reliably by assuming that records are reliable. Such an inference is backed by the assumption that the recording device was "ready" at a time before the event, and there was another recording device measuring the "ready" state of that one, and so on. As discussed before, the records arrow is explained by PH. Moreover, if the epistemic arrow has physical necessity (or high objective probability), as is required to avoid epistemic self-undermining, whatever underlies it also has physical necessity. In a way similar to the records arrow, the epistemic arrow has to be non-accidental, otherwise our beliefs about past experiments would have a low probability of being accurate.

\textit{The counterfactual arrow.} I am at home right now. If I had been in my office, the future would be somewhat different but the past would have been pretty much the same. Trump did not press the nuclear button on Independence Day this year. If he had,  events on Labor Day would be dramatically different but events on Memorial Day would have been pretty much the same. Why is there a temporal asymmetry of counterfactual dependence? The semantics for counterfactuals is a controversial issue. It is not clear if there is a unified theory that explains every instance of counterfactuals in ordinary language. But if we focus on the counterfactuals that are important for control, decision, and action,  it is often accepted that such counterfactuals are backed by laws. This is made explicit in \cite{LewisCDTA}'s metric for comparing similarity relations among worlds, but it should also be compatible with a strict-conditional approach. If the laws are temporally asymmetric, and if laws entail that changes in the present macrostate would lead to vast differences in the future but not much in the past, then the counterfactual arrow has an explanation. 

However, given equations (\ref{HE}) in classical mechanics or equations (\ref{SE}, \ref{GE}) in unitary quantum mechanics, changes to the current state (such as the location of Trump's index finger and the location of the nuclear button) will lead to macroscopic differences in both directions of time. It is only by assuming PH as a fundamental law can we explain the following: most physically possible trajectories compatible with the current macrostate will be such that if Trump had pressed the button, most future trajectories would be macroscopically different from the actual ones but the past trajectories would be pretty much the same. This is also due to the records arrow. Assuming PH, there will be an abundance of records  about the past. In so far as PH makes it very likely that those records are reliable, PH constrains  the past histories of the trajectories even if certain macroscopic features get changed in the present state. However, few, if any,  records exist about the future, so there is no such constraint on future macrostates. 

The correct counterfactual semantics will no doubt involve context sensitivity and other parameters. Still,  it is hard to deny that laws of nature play an important role in determining the truth values of counterfactuals. 

\textit{The intervention arrow.} We can exert influence towards future events but we can no longer act to bring about changes to the past. The intervention arrow is intimately connected to the counterfactual arrow, and it is not clear which is conceptually prior. Some contemporary analyses of influence and intervention are couched in the causal modeling framework. In that framework, we  have directed acyclic graphs with variables representing events and arrows representing the direction of effect. But if the fundamental dynamical laws are time-symmetric, what is the scientific explanation for these arrows? Often, the arrows are taken to be primitives in the causal model, left unexplained. However, if PH explains the counterfactual arrow, then it can also explain the arrow of intervention.  We can flesh out the language of intervention in terms of intervention counterfactuals, and the arrow of intervention counterfactuals can be explained in a similar way by the records arrow and PH. \cite{LoewerCatSLaw} provides such an account.

The arguments from the entropic arrow (the Second Law) and the other arrows can be taken together as an inference to the best explanation. PH (and SP) ground these asymmetries of time. Moreover, to explain them satisfactorily, we need to postulate PH as a fundamental law and we need SP to provide objective probabilities. An opponent may take all of these arrows to be fundamental features of the world, and they can postulate them as primitives in the theory. But that would strike many as a fragmented and unsatisfactory view. Postulating PH and SP in addition to the dynamical laws is a much simpler and more unified way to think about the various arrows of time: from a set of simple axioms, we can derive all of the temporally asymmetric regularities---we get a big bang for the buck.

\subsection{Arguments from Metaphysical Accounts of Laws}

Humeanism provides a natural home for PH to be a fundamental law and for SP to specify objective probabilities. According to \cite{LewisNWTU}, the fundamental laws and postulates of objective probabilities are the \textit{axioms} of the best system that are true about the mosaic and optimally balances various theoretical virtues such as simplicity, informativeness, and fit. On this account, the dynamical equations such as equations (\ref{HE}, \ref{SE}, and \ref{GE}) could be axioms of their respective best systems and the GRW chances could be the objective probabilities in a GRW world. Can the classical Mentaculus, quantum Mentaculus, and the Wentaculus count as axiomatizations of the best system? This depends on whether PH can count as a fundamental Lewisian law and whether SP can count as objective probabilities on the best-system account. Anticipating the need to add a boundary condition into the best system, \cite{LewisNWTU} writes,

\begin{quotation}
  A law is any regularity that earns inclusion in the ideal system. (Or, in case of ties, in every ideal system.) The ideal system need not consist entirely of regularities; particular facts may gain entry if they contribute enough to collective simplicity and strength. (For instance, certain particular facts about the Big Bang might be strong candidates.) But only the regularities of the system are to count as laws. (p.367)
\end{quotation}
 But if a statement such as PH is axiomatic in the best system, why not count it towards laws? In the same paper (p. 368), Lewis distinguishes between fundamental laws and derived laws. He suggests that fundamental laws are those statements that the ideal system takes as \textit{axiomatic} and invokes only perfectly natural properties. But PH certainly is axiomatic in the Mentaculus and the Wentaculus. Moreover, PH can be stated in the fundamental language of the respective theory. (There will be some residual vagueness, which we discuss in \S4.3.) So it seems that Lewis should be open to the idea that PH is a fundamental law according to the best-system account.\footnote{For a related point,  see \citep{callender2004measures}. }
 
 Hence, if we are committed to the Humean conception that laws supervene on the mosaic in the way specified by the best-system account,  we are led to accept the fundamental nomic status of PH.  (Similarly, as \cite{loewer2001determinism} argues, Lewis's version of Humeanism also has room to recognize the probabilities specified by SP as objective.) 
 
 On the Humean theory, given facts about the mosaic, the theoretical virtues \textit{metaphysically determine} what the laws are. They are constitutive of laws. Laws are just certain ideal summaries of facts in the world. Laws are nothing over and above the mosaic. 
 
 On  non-Humean theories, however, laws do not supervene on the mosaic. Laws may be as fundamental as the mosaic itself. Following \cite{hildebrand2013can}, we can distinguish between  two types of non-Humean theories: 
 \begin{enumerate}
  \item Primitivism: fundamental laws are primitive facts in the world. 
  \item Reductionism: fundamental laws are analyzed in terms of something outside the mosaic. 
\end{enumerate}
\cite{carroll1994laws} and \cite{MaudlinMWP} maintain  primitivist versions of non-Humeanism. \cite{hildebrand2013can} provides a survey of the reductionist versions according to which laws are further explained by relations among universals, dispositions, essences, or some other more fundamental entities. It is not clear what the further analysis buys us. It is not clear to me how to reformulate various modern physical laws and objective probabilities in terms of those entities, and it is less clear to me what advantages there are to reduce laws to something further.\footnote{\cite{hildebrand2013can} suggests that reductionist theories have an advantage over primitivist theories for answering the problem of induction. I disagree, but I set it aside for future work.} Here I agree with Maudlin that the concept of laws seems more familiar to us than the  concepts employed in the further analysis (such as in terms of dispositions, universals, and the like). Maudlin's version of primitivism is influential in contemporary discussions of the metaphysics of laws in philosophy of physics. However, Maudlin (2007) favors a more restrictive version of primitivism that I call \textit{Dynamical Law Primitivism}: 
\begin{description}
  \item[Dynamical Law Primitivism] Fundamental laws are primitive facts in the world, and only dynamical laws can be fundamental laws. 
\end{description}
This view is connected to Maudlin's view about the intrinsic and primitive arrow of time. In contrast, the spirit of the present project is to analyze time's arrow in terms of something else. In fact, the extra commitment about the primitive arrow of time  can be disentangled from the basic idea about how laws govern. 

The basic non-Humean idea is simply that laws really \textit{govern}. They metaphysically explain why the nomological possibilities are the way they are and why things are as constrained as they are. The metaphysical explanation can take the form of  constraints: given $S(t_0)$, some complete specification of the state of the universe at some time, there is a constraint on what the history of the world is like. If the theory is deterministic, then there is only one microscopic history compatible with the $S(t_0)$. A fundamental dynamical law is a kind of conditional constraint. Constraints can take other forms, such as by selecting a space of possible histories. This is the form of certain equations in general relativity and Maxwellian electrodynamics. It is also true of PH, as it  selects  ``a set of admissible histories among the solutions of the dynamical laws.'' There is conceptual space for a minimalist conception of primitivism that places no restriction on the form of fundamental laws and in particular, not all of them have to be dynamical laws. The basic view is this: 
\begin{description}
  \item[Minimal Primitivism] Fundamental laws are primitive facts in the world; there is no restriction on the form of fundamental laws. In particular, boundary conditions can be fundamental laws. 
\end{description}
Even though fundamental laws can take on any form, we expect them to be relatively simple and informative. These theoretical virtues are no longer \textit{constitutive} of what laws are, but they can serve as  our best guides to find  the primitive laws\footnote{ Why is this expectation rational? I do not know of any non-circular justification. We can appeal to the success of physics  and the discovery of simple and informative laws in the past. We may also appeal to a deeper ``meta-law'' that metaphysically constrain the physical laws.}:
\begin{description}
  \item[Epistemic Guides] Even though theoretical virtues such as simplicity and informativeness are not constitutive of fundamental laws, they are good epistemic guides for discovering the fundamental laws. 
\end{description}
\cite{chenandgoldstein} develop this idea in more detail.  It seems to me that Minimal Primitivism is a good version of non-Humeanism, and it may well be one that best fits our scientific practice and the actual conception of laws. The minimal primitivist view does not commit to an intrinsic and irreducible arrow of time, making it compatible with the Boltzmannian project of analyzing time's arrow in terms of the entropy gradient.  PH, as we have discussed already, is virtuous in the right ways. It provides a simple explanation for the restrictions of physical possibilities and the overwhelming probability of irreversibility. According to the Epistemic Guides on the Minimal Primitivist conception, we have found strong evidence that  PH is a candidate fundamental law. 

Hence, both  Humeanism and  non-Humeanism (in the minimalist form) support the idea that PH is a candidate fundamental law.

\subsection{New Light on Quantum Entanglement}
I suggest that there is a new reason to take PH as a fundamental law: it can help us solve a  long-standing puzzle in the foundations of quantum mechanics. One of the chief innovations of quantum theory that has no classical analog is quantum entanglement. It is also the origin of the quantum measurement problem. If we solve the  measurement problem using one of the three strategies discussed in \S2: along the lines of Everett, Bohm, and GRW, we are still left with the quantum state that plays an important dynamical role in the respective theories.  Hence, the puzzle of quantum entanglement can be traced to the nature of the quantum state. What does the quantum state represent physically? What is it in the world?  Given the role of $\Psi$ in formulating well-posed initial-value problems in EQM and BM, and its role in dynamical collapses in GRW, it is reasonable to think that $\Psi$ represents something objective. Here are some options for a realist interpretation (see \citep{chen2019realism} for more detail): 
\begin{enumerate}
  \item High-dimensional field: on this view, the fundamental space is isomorphic to the 3N-dimensional configuration space; the wave function represents a physical field on that space \citep{AlbertEQM}. Even  the defenders acknowledge that this view has highly counter-intuitive consequences. It is also an open question whether it really succeeds in recovering the manifest image of lower-dimensional objects.
  \item Low-dimensional multi-field: on this view, the fundamental space is the physical space(time); the wave function represents a ``multi-field'' that assigns physical quantities to every spatial region composed of $N$ points \citep{forrest1988quantum, belot2012quantum, ChenOurFund, Hubert2018}. However, this view also appears to have undesirable consequences. In the multi-field interpretation of Everettianism, since entanglement relations are still in the 4-dimensional mosaic, its Lorentz-invariance comes at a surprising cost---the failure of what \cite{albert2015after} calls \textit{narratability}. This also arises in \cite{wallace2010quantum}'s spacetime state realist interpretation of Everett. For the Bohmian framework, the multi-field guides particles, but there is no influence (back-reaction) of the particles on the multi-field, even though both the particles and the multi-field are fundamental material entities.  
  \item Nomological interpretation: on this view, the fundamental space is the physical space(time) and the fundamental ontology consists in particles, fields, or flashes on that space; the wave function represents a physical law \citep{goldstein1996bohmian}, \citep{goldstein2001quantum} and \citep{goldstein2013reality}. Given the problems of the other two approaches, the nomological interpretation remains a promising solution.  However, it  faces problems of a different kind. First, $\Psi_t$ is time-dependent. Can nomological entities change in time? I do not see why not. Moreover, as defenders of this view have long recognized, if the universal quantum state obeys the Wheeler-DeWitt equation $H\Psi=0$ then $\Psi$ will be time-independent and not changing. Second, the universal wave function is a very detailed function and may be too complicated to be a law. This problem seems much more serious. Call this the problem of complexity.
\end{enumerate}

Taking PH to be a fundamental law provides a solution to the problem of complexity in the nomological interpretation of the quantum state. This solution works in the Wentaculus framework, where IPH is given a fundamental nomic status.  For IPH, the normalized projection onto $\mathscr{H}_{PH}$ contains no more and no less information than $\mathscr{H}_{PH}$, specified by PH in the Mentaculus. If  $\mathscr{H}_{PH}$ is simple and informative enough to be nomological, then so is its normalized projection, which is $W_{IPH}(q,q', t_0)$. That is, we can afford the same status of a fundamental law to   $W_{IPH}(q,q', t_0)$. In the Everettian Wentaculus, we can interpret $W_{IPH}(q, q', t_0)$ as a law that determines the ``local beables,'' such as a matter-density  field in physical space. In the Bohmian Wentaculus, we can interpret $W_{IPH}(q,q', t_0)$ as similar to the Hamiltonian function $H(p,q)$: providing a velocity field of particle trajectories.  In the GRW Wentaculus, we can interpret $W_{IPH}(q,q', t_0)$ as providing  conditional probabilities for the configurations of``local beables,'' such as a matter-density field or flashes in spacetime. 

However, to solve the complexity problem, it is not sufficient for IPH to be a contingent initial condition; it is crucial that IPH has the status of a fundamental law. If it is nomologically possible for $W_0$ to differ from the state specified by IPH, then the initial quantum state could well be (i.e. physically possible) too complicated to be regarded as nomological. Hence,  only by assuming the nomic status of IPH do we obtain a  solution to the problem of complexity, thereby arriving at a satisfactory way to understand the nature of the quantum state. In contrast to the first two interpretations according to which quantum entanglement relations are in the material ontology, the nomological interpretation of the quantum state locates the origin of entanglement in the laws. And by taking a nomological interpretation of the quantum state in the Wentaculus framework, we see a unified solution to  two problems in the foundations of physics: the problem of irreversibility and the nature of the quantum state. Again, this is compatible with both Humeanism and (the minimal form of) non-Humeanism about laws. (For more detail, see \citep{chen2018IPH, chen2018HU}.)

\section{Apparent Conflicts with Our Concept of Laws of Nature}
In the previous section, I have provided positive arguments for the fundamental, nomic status of PH. In this section, I discuss three apparent conflicts between these arguments and the ordinary conceptions of laws of nature. These apparent conflicts may explain some people's hesitation in accepting the fundamental, nomic status of PH. However, as I argue, these conflicts are merely apparent if we adopt the Mentaculus framework, and at any rate, they become even less worrisome in the Wentaculus framework. 

\subsection{Boundary Condition Laws?}
It is often said that PH is merely a boundary condition. The contrast between boundary conditions and fundamental laws can be seen in the differences between  the paradigm cases of dynamical laws---such as equations (\ref{HE}, \ref{SE},  \ref{GE}, \ref{VNM}, and \ref{WGE}), and  boundary conditions that select  subclasses of the dynamically possible trajectories. Boundary conditions do not \textit{directly} play a dynamical role. 

It is not clear how to make this worry precise. First, it is unclear why every law has to be dynamical. Second,  by restricting the possible initial conditions, a boundary-condition law can be an important ingredient in the theory, as in the case of the Mentaculus. In fact, in the Wentaculus framework, the boundary condition IPH plays a \textit{direct} dynamical role, akin to the dynamical role of the Hamiltonian function in classical mechanics. For example, in the Bohmian version,  IPH (\ref{IPH}),  the von Neuman equation (\ref{VNM}), and the guidance equation (\ref{WGE})  can be combined into one equation: 
\begin{equation}\label{N4}
\frac{dQ_i}{dt} = 
\frac{\hbar}{m_i} \text{Im} \frac{\nabla_{q_{i}}   W_{IPH} (q, q', t)}{W_{IPH} (q, q', t)} (Q) = 
\frac{\hbar}{m_i} \text{Im} \frac{\nabla_{q_{i}}   \bra{q} e^{-i \hat{H} t/\hbar} \hat{W}_{IPH} ( t_0) e^{i \hat{H} t/\hbar} \ket{q'} }{ \bra{q} e^{-i \hat{H} t/\hbar} \hat{W}_{IPH} (t_0) e^{i \hat{H} t/\hbar} \ket{q'}} (q=q'=Q)
\end{equation}
Hence, in the Bohmian version, IPH does not just select a subclass of velocity fields;  it pins down a unique velocity field. In the Everettian version, IPH does not just select a subclass of possible multiverses; it pins down a unique one. In the GRW version, IPH is directly involved in setting the chances of collapses. 

There is a related worry about admitting boundary condition laws. Typically we distinguish between dynamical laws and initial conditions. If initial conditions can be laws, then how can we distinguish between laws and contingent data? It seems that the distinction would collapse. However, that is not the case. Some boundary conditions such as PH have the elite status as fundamental laws, but it does not follow that all boundary conditions are similarly elite.  This is because not all boundary conditions exemplify the optimal balance of required theoretical virtues (either as constitutive of what laws are or as epistemic guides to primitive laws), including simplicity, informativeness, and fit.


One may  worry that it is odd to have a fundamental law that refers to a particular time ($t_0$). Our most familiar fundamental laws are  general statements that do not refer to a particular time or place. But why is it a requirement that laws cannot refer to particular events? Suppose some places or times are in fact physically distinguished; then it seems appropriate for laws to refer to them. Think about the Aristotelian tendency for things to move towards the center of the (geocentric) universe. If that is indeed the case, then we should have a fundamental law describing that motion, and a simple candidate would just be to state that the center of the world, $C_0$, is the place towards which things evolve. Similarly, if the initial time, $t_0$, is indeed special (as the initial state accounts for a great many regularities), then it is appropriate to postulate a law that refers to $t_0$.

\subsection{Non-Dynamical Chances?}
Another worry concerns the nature of objective probabilities. The Mentaculus postulates both PH and SP. PH and SP  share the explanatory burden, so they should have the same status. It is important that the probabilities of SP be objective. However, if the dynamics are deterministic (as in the Bohmian and the Everettian versions), objective probabilities seem to be either 0 or 1. How can non-trivial probabilities be objective and represent something beyond subjective credences?  (In so far as typicality plays a similar  explanatory role as probability, TP may face the same \emph{prima facie} problem as SP.) 

Humeanism has the resources to solve this problem. \cite{loewer2001determinism} suggests that deterministic ``chances'' can gain entry in the Lewisian best system by the informativeness they bring and the simplicity of the postulate, such as the uniform measure specified by SP. On non-Humeanism, how to understand deterministic ``chances''  is an open problem. On Minimal Primitivism, perhaps the notion of primitive constraining (of the initial state) can come in degrees that can be represented by probabilities. Another way to allow deterministic ``chances" is to use the notion of typicality specified by TP, which can be interpreted as only allowing the initial conditions that are \textit{typical}. On this understanding, abnormal, anti-entropic initial conditions are not nomologically possible. (A similar  problem arises in Bohmian mechanics, which requires a quantum equilibrium distribution in addition to the deterministic dynamical laws, to deduce the usual Born rule for subsystems.)

In the Wentaculus framework,  it is clear that there is an objective anchor for SP (and TP). Since IPH selects a unique initial quantum state of the universe, we no longer need a probability distribution over initial quantum states. However, as a purely mathematical fact, the $W_{IPH}(t_0)$ induces a ``uniform'' probability distribution over pure states. This is not a fundamental postulate of probability in the theory, unlike SP or TP in the Mentaculus. It arises as a mathematical consequence of the objective quantum microstate of the universe. This emergent probability distribution, though not fundamental, can play the same role in statistical analysis about typical behaviors. For example, the usual conjecture about typical pure states approaching equilibrium can be translated into the following: most parts of the density matrix will be very close to the equilibrium subspace, which is equivalent to the claim that the density matrix will approach equilibrium. Hence, the Wentaculus framework provides an objective anchor  for SP and TP.\footnote{However, it does not answer the related question about the nature of the quantum equilibrium distribution in Bohmian theories. But that is different from the issue of the status of SP and TP.}

\subsection{Nomic Vagueness?}

The final worry about the fundamental nomic status of PH has to do with the fact that PH is vague, and an exact version of PH would be arbitrary in an unprecedented way. In Figure 1 of \S2.1, I made clear that the boundaries of macrostates are fuzzy, and the macrostates only form an exact partition if we stipulate some arbitrary choices of the parameters for coarse-graining. These are the C-parameters: the size of coarse-graining cells, the exact correspondence between macroscopic quantities and functions on phase space, and (in the quantum case) the cut-off value for macrostate inclusion. There are better or worse ways to choose the C-parameters. But it is implausible that there are some exact values of C-parameters as known to Nature. The vagueness  comes up in the Classical Mentaculus and the Quantum Mentaculus in how  PH selects an initial macrostate that constrains the initial microstate. In the quantum case, PH only selects an exact subspace in Hilbert space when we choose some arbitrary C-parameters. 

In the Quantum Mentaculus, can we stipulate that there is an exact subspace $\mathscr{H}_{PH}$ as known to Nature? This leads to what I call \textit{untraceable arbitrariness}. There is an infinity of \textit{admissible} changes\footnote{Admissibility here is  vague and rightly so. It can be interpreted as some measure of simplicity of theories. We want PH to be simple enough, and different ways of carving out the boundary will lead to different exact versions of PH. But we only want to consider those versions that are sufficiently simple (and not too gerrymandered).} to the boundary of  $\mathscr{H}_{PH}$ that do not change the nomological status of most microstates compatible with  $\mathscr{H}_{PH}$. This is  unlike the kind of arbitrariness of natural constants or other fundamental laws. For example, any change to the value of the gravitational constant in Newtonian theory will make most worlds (compatible with the original Newtonian theory) impossible. What about the case of stochastic theories? Are the dynamical chances traceable? Yes they are, but not in the way of changing status from physically possible to physically impossible. The traceable changes are reflected in the probabilistic likelihood of most worlds. 

 \begin{figure}
\centerline{\includegraphics[scale=0.27]{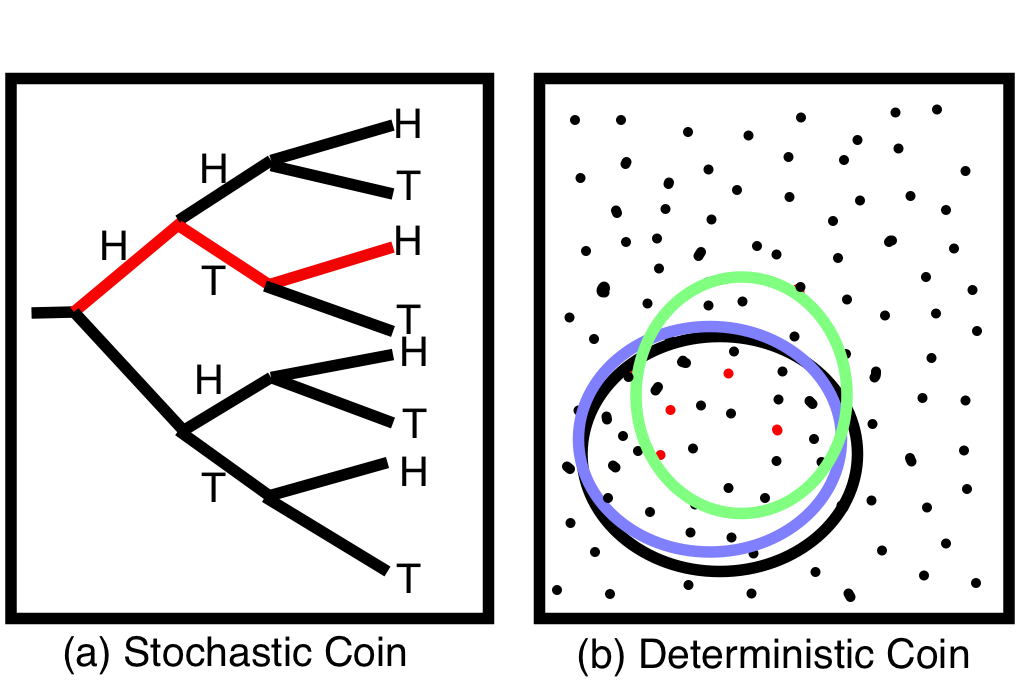}}
\caption{Illustration of nomic vagueness using the contrast between a stochastic coin and a deterministic coin. What is actually observed is this sequence: Heads, Tails, Heads.  }
\end{figure}

I discuss this in more details in \citep{chen2018NV}. Here I provide a simple illustration by considering mechanisms for flipping a coin (see Figure 2(a)). Suppose we have a stochastic coin and it is flipped three times. It landed Heads, Tails, and Heads. In the color version of the diagram, the possible sequences are marked in black and the actual sequence is marked in red. The simplest chance hypotheses are going to be:
\begin{itemize}
  \item $H_{\alpha}$: the chance of landing Heads at each flip is the same, and it is $\alpha$. 
\end{itemize}
So, this is a one-parameter family of chance hypotheses. $H_{0.5}$ is the hypothesis that the coin is fair, $H_{1}$ is the hypothesis that the coin always lands Heads, and so on. Among these hypotheses, a sequence of coin flips will select exactly one of the chance hypotheses from this class based on which one receives the highest likelihood value.  In this case, it is $H_{2/3}$ with the highest likelihood value being $4/27$. Conversely, $H_{2/3}$ will assign a determinate chance to every sequence of coin flips. If Nature stochastically acts according to $H_{2/3}$, then changing the chance of Heads even slightly, say to $0.65$, will change the chance of every sequence.  So, it is in this sense that stochastic theories are traceable. Of course, we can imagine a more gerrymandered chance hypothesis according to which half of the time, the coin operates with a $5/6$ chance of landing Heads, and the other half of the time, the coin operates with a $3/6$ chance of landing Heads. This will predict the same chance as $H_{2/3}$ to every sequence, but it is far less simple. The gerrymandered chance hypothesis is not a serious competitor to $H_{2/3}$. The actual hypothesis is \textit{by far} simpler and more fit than any competitor.

Traceability is lost in the case of the deterministic coin. In this case, there are no dynamical transition chances. The objective probabilities come from probabilities over initial conditions. Suppose a deterministic coin is flipped and it lands Heads, Tails, and Heads. In the color version of Figure 2(b), the red dots represent the initial conditions of the coin (which also include details about the flipping mechanism) that deterministically lead to the sequence HTH, and the black dots represent initial conditions that lead to other outcomes (such as HHH, TTT, and so on). To simplify things, suppose the sample space is finite, so there are only finitely many initial conditions to consider. Then we can draw different probabilistic hypotheses as different ``circles'' over initial conditions. Suppose the Black Circle encloses 4 red dots and 23 black dots. Then it represents a  probabilistic hypothesis that all and only the dots within the Black Circle are possible initial conditions and each dot has equal probability. The Black Circle has the highest likelihood given the data of HTH. If the Green Circle encloses 4 red dots and 25 black dots, then the Green Circle is less likely than the Black Circle given HTH. However, it is easy to have a nearby circle, say the Blue Circle, that (like the Black Circle) has 4 red dots and 23 black dots. This is possible if the red dots are sufficiently localized in state space such that it is easy to preserve their proportion to the black dots while changing the boundary of the circle. 

Moreover, specifying the Blue Circle need not be more complicated than specifying the Black Circle. They are the same kind of probabilistic hypotheses, and there is no reason to think that one is more gerrymandered than the other, unlike the situation with the stochastic coin where to recover the same likelihood, one has to resort to time-dependent chances. This reasoning can be generalized if the state space gets richer and the sequence of coin flips gets longer. There may be infinitely many ways to slightly change the boundary of the Black Circle and keep the relative proportions \textit{constant}. This means that there will be a large class of probabilistic hypotheses that have the same likelihood given a particular history of coin flips. No particular hypothesis is far simpler and more fit than all competitors. Hence, super-empirical virtues such as simplicity will become more relevant. Furthermore, since the variation of the boundary is incremental, comparing simplicity can generate a sorites series: is there a determinate class of hypotheses that are simple enough? It is implausible for there to be a sharp line. Hence, we have a vague ``collection'' of hypotheses that pass the simplicity bar.  

The contrast between the stochastic coin and the deterministic coin is  analogous to the comparison between GRW theories and a vague law such as PH. GRW chances are traceable, but the boundary of the PH macrostate and the exact probability distribution are untraceable. Hence, there are reasons to think that if PH is a fundamental law in the Mentaculus, then it is a vague law. I call this phenomenon \textit{nomic vagueness}. 

However, it is not clear why vagueness disqualifies a statement from being a fundamental law. After all, we should be led by empirical evidence and scientific practice to consider what the laws are, and our metaphysical commitments to precision and exactness should not be given absolute priority. It is a surprising consequence that a fundamental law can be vague. This gives us reason to think that perhaps the final theory of the world will not be completely mathematical expressible, in so far as vagueness and higher order vagueness defy classical logic and set-theoretic mathematics. This is a radical consequence about nomic vagueness that deserves more attention \citep{ChenNS2020}. 

Nevertheless, the situation is different in the Wentaculus. It gets rid of nomic vagueness without introducing untraceable arbitrariness.  According to IPH, the initial macrostate and the initial microstate is represented by the quantum state of the universe---$W_{IPH}(t_0)$. It enters directly into the fundamental micro-dynamics. Hence, $W_{IPH}(t_0)$ will be traceable from the perspective of two realist interpretations of the quantum state \citep{chen2019realism}:
\begin{enumerate}
	\item $W_{IPH}(t_0)$ is ontological: if the initial density matrix represents something in the fundamental material ontology, IPH is obviously traceable. Any changes to the physical values $W_{IPH}(t_0)$ will leave a trace in every world compatible with IPH. 
	\item $W_{IPH}(t_0)$ is nomological: if the initial density matrix is on a par with the fundamental laws, then $W_{IPH}(t_0)$ plays the same role as the classical Hamiltonian function or fundamental dynamical constant of nature. It is traceable in the Everettian version with a matter-density ontology as the initial matter-density is obtained from $W_{IPH}(t_0)$. It is similarly traceable in the GRW version with a matter-density ontology. For the GRW version with a flash ontology, different choices of $W_{IPH}(t_0)$ will, in general, lead to different probabilities of possible macro-histories. In the Bohmian version, different choices of $W_{IPH}(t_0)$ will lead to different velocity fields such that for typical initial particle configurations (and hence typical worlds compatible with the theory), they will take on different trajectories.   
 \end{enumerate}
The traceability of $W_{IPH}(t_0)$ is due to the fact  that we have connected the low-entropy macrostate (now represented by $W_{IPH}(t_0)$) to the micro-dynamics (where  $W_{IPH}(t_0)$ appears). Hence, $W_{IPH}(t_0)$ is playing a dual role at $t_0$ (and only at that time): it is both the microstate and the macrostate. In contrast, the untraceability of $\Gamma_0$ in classical mechanics is due to the fact that  classical equations of motion directly involve only the microstate $X_0$, not $\Gamma_0$. Similarly, $\mathscr{H}_{PH}$ in the standard wave-function formulation  is untraceable because the Schr\"odinger equation directly involves only the wave function, not $\mathscr{H}_{PH}$. Many changes could be made to $\Gamma_0$ and $\mathscr{H}_{PH}$ that would not trickle down whatsoever in typical worlds compatible with these postulates. The Mentaculus, but not the Wentaculus, faces a dilemma between nomic vagueness and untraceable arbitrariness.

\section{Conclusion}

I have argued that, in the Boltzmannian framework, PH is a candidate fundamental law of nature. Such a view  is supported by the theoretical roles PH plays in the theory. In arguing for the Nomic Status and the Axiomatic Status of PH, we see that whether it is a law makes a difference to many other issues in the foundations of physics. Moreover, its nomic status calls for some re-thinking about the nature of physical laws. I suggest that, according to Humeanism and a minimal version of non-Humeanism, boundary conditions can be fundamental laws,  SP and TP can be objective, and fundamental laws and chances can be vague. The conflicts with our concept of laws of nature are merely apparent, and in any case, they become much less worrisome if we adopt the Wentaculus framework.  Hence,  the view that PH is a candidate fundamental law  should be more widely accepted than it is now.


\section*{Acknowledgement}
My ideas in this paper have been influenced by discussions with many people over the years. I am especially grateful to David Albert, Craig Callender, Sheldon Goldstein,  Barry Loewer, and Roderich Tumulka. I would also like to thank  
Eugene Chua,
Saakshi Dulani,
Veronica Gomez,
Ned Hall,
Tim Maudlin,
Kerry McKenzie,
Elizabeth Miller,
Jill North,
Charles Sebens,
Ted Sider,
Cristi Stoica,
Anncy Thresher,
David Wallace,
Brad Weslake, 
Isaac Wilhelm,
Eric Winsberg, 
and the participants in my graduate seminar on the arrows of time at UCSD in  spring  2020.


\bibliography{test}

\begin{thebibliography}{}

\bibitem[\protect\citename{Albert, }1996]{AlbertEQM}
Albert, David~Z. 1996.
\newblock Elementary Quantum Metaphysics.
\newblock {\em Pages  277--84 of:} Cushing, J.~T., Fine, A., \& Goldstein, S.
  (eds), {\em Bohmian Mechanics and Quantum Theory: An Appraisal}.
\newblock Dordrecht: Kluwer Academic Publishers.

\bibitem[\protect\citename{Albert, }2000]{albert2000time}
Albert, David~Z. 2000.
\newblock {\em Time and chance}.
\newblock Cambridge: Harvard University Press.

\bibitem[\protect\citename{Albert, }2015]{albert2015after}
Albert, David~Z. 2015.
\newblock {\em After physics}.
\newblock Cambridge: Harvard University Press.

\bibitem[\protect\citename{Allori {\em et~al.\ }\relax,
  }2013]{allori2013predictions}
Allori, Valia, Goldstein, Sheldon, Tumulka, Roderich, \& Zangh{\`\i}, Nino.
  2013.
\newblock Predictions and primitive ontology in quantum foundations: a study of
  examples.
\newblock {\em The British Journal for the Philosophy of Science}, {\bf 65}(2),
  323--352.

\bibitem[\protect\citename{Ashtekar \& Gupt, }2016]{ashtekar2016initial}
Ashtekar, Abhay, \& Gupt, Brajesh. 2016.
\newblock Initial conditions for cosmological perturbations.
\newblock {\em Classical and Quantum Gravity}, {\bf 34}(3), 035004.

\bibitem[\protect\citename{Barrett, }1996]{barrett1996empirical}
Barrett, Jeffrey~A. 1996.
\newblock Empirical adequacy and the availability of reliable records in
  quantum mechanics.
\newblock {\em Philosophy of Science}, {\bf 63}(1), 49--64.

\bibitem[\protect\citename{Bell, }1980]{bell1980broglie}
Bell, John~S. 1980.
\newblock De {B}roglie-{B}ohm, delayed-choice, double-slit experiment, and
  density matrix.
\newblock {\em International Journal of Quantum Chemistry}, {\bf 18}(S14),
  155--159.

\bibitem[\protect\citename{Belot, }2012]{belot2012quantum}
Belot, Gordon. 2012.
\newblock Quantum states for primitive ontologists.
\newblock {\em European Journal for Philosophy of Science}, {\bf 2}(1), 67--83.

\bibitem[\protect\citename{Boltzmann, }1964]{boltzmann2012lectures}
Boltzmann, Ludwig. 1964.
\newblock {\em Lectures on gas theory}.
\newblock Berkeley: University of California Press.

\bibitem[\protect\citename{Callender, }2004]{callender2004measures}
Callender, Craig. 2004.
\newblock Measures, explanations and the past: Should `special' initial
  conditions be explained?
\newblock {\em The British journal for the philosophy of science}, {\bf 55}(2),
  195--217.

\bibitem[\protect\citename{Callender, }2011]{sep-time-thermo}
Callender, Craig. 2011.
\newblock Thermodynamic Asymmetry in Time.
\newblock {\em In:} Zalta, Edward~N. (ed), {\em The Stanford Encyclopedia of
  Philosophy}, fall 2011 edn.
\newblock Metaphysics Research Lab, Stanford University.

\bibitem[\protect\citename{Carroll, }1994]{carroll1994laws}
Carroll, John~W. 1994.
\newblock {\em Laws of nature}.
\newblock Cambridge University Press.

\bibitem[\protect\citename{Carroll, }2010]{carroll2010eternity}
Carroll, Sean. 2010.
\newblock {\em From eternity to here: the quest for the ultimate theory of
  time}.
\newblock Penguin.

\bibitem[\protect\citename{Carroll \& Chen, }2004]{carroll2004spontaneous}
Carroll, Sean~M, \& Chen, Jennifer. 2004.
\newblock Spontaneous Inflation and the Origin of the Arrow of Time.
\newblock {\em arXiv preprint hep-th/0410270}.

\bibitem[\protect\citename{Chen, }2017]{ChenOurFund}
Chen, Eddy~Keming. 2017.
\newblock Our Fundamental Physical Space: An Essay on the Metaphysics of the
  Wave Function.
\newblock {\em Journal of Philosophy}, {\bf 114: 7}.

\bibitem[\protect\citename{Chen, }2018]{chen2018IPH}
Chen, Eddy~Keming. 2018.
\newblock Quantum Mechanics in a Time-Asymmetric Universe: On the Nature of the
  Initial Quantum State.
\newblock {\em The British Journal for the Philosophy of Science}, {\bf
  forthcoming}.

\bibitem[\protect\citename{Chen, }2019a]{chen2019quantum1}
Chen, Eddy~Keming. 2019a.
\newblock Quantum States of a Time-Asymmetric Universe: Wave Function, Density
  Matrix, and Empirical Equivalence.
\newblock {\em arXiv:1901.08053}.

\bibitem[\protect\citename{Chen, }2019b]{chen2019realism}
Chen, Eddy~Keming. 2019b.
\newblock Realism about the wave function.
\newblock {\em Philosophy Compass}, {\bf 14}(7).

\bibitem[\protect\citename{Chen, }2020a]{chen2018HU}
Chen, Eddy~Keming. 2020a.
\newblock From Time Asymmetry to Quantum Entanglement: A {H}umean Unification.
\newblock {\em No\^us}, {\bf forthcoming}.

\bibitem[\protect\citename{Chen, }2020b]{chen2018NV}
Chen, Eddy~Keming. 2020b.
\newblock Nomic Vagueness.
\newblock {\em arXiv preprint arXiv:2006.05298}.

\bibitem[\protect\citename{Chen, }2020c]{chen2020time}
Chen, Eddy~Keming. 2020c.
\newblock Time's Arrow and De Se Probabilities.
\newblock {\em arXiv preprint arXiv:2001.09972}.

\bibitem[\protect\citename{Chen, }2020d]{chen2018valia}
Chen, Eddy~Keming. 2020d.
\newblock Time's Arrow in a Quantum Universe: On the Status of Statistical
  Mechanical Probabilities.
\newblock {\em In:} Allori, Valid (ed), {\em Statistical Mechanics and
  Scientific Explanation: Determinism, Indeterminism and Laws of Nature}.
\newblock Singapore: World Scientific.

\bibitem[\protect\citename{Chen, }2020e]{ChenNS2020}
Chen, Eddy~Keming. 2020e.
\newblock Welcome to the Fuzzy-Verse.
\newblock {\em New Scientist}, {\bf 247}(3298), 36--40.

\bibitem[\protect\citename{Chen \& Goldstein, }2020]{chenandgoldstein}
Chen, Eddy~Keming, \& Goldstein, Sheldon. 2020.
\newblock Minimal Primitivism about Laws of Nature.
\newblock {\em In preparation}.

\bibitem[\protect\citename{Chen \& Tumulka, }2020]{chen2020uniform}
Chen, Eddy~Keming, \& Tumulka, Roderich. 2020.
\newblock Uniform Probability Distribution Over All Density Matrices.
\newblock {\em arXiv preprint arXiv:2003.13087}.

\bibitem[\protect\citename{Demarest, }2019]{demarest2019mentaculus}
Demarest, Heather. 2019.
\newblock Mentaculus Laws and Metaphysics.
\newblock {\em Principia: an international journal of epistemology}, {\bf
  23}(3), 387--399.

\bibitem[\protect\citename{D{\"u}rr {\em et~al.\ }\relax,
  }1996]{goldstein1996bohmian}
D{\"u}rr, Detlef, Goldstein, S, \& Zangh\`i, N. 1996.
\newblock Bohmian Mechanics and the Meaning of the Wave Function.
\newblock {\em In:} {\em in Experimental Metaphysics: Quantum Mechanical
  Studies in honor of Abner Shimony}.

\bibitem[\protect\citename{D{\"u}rr {\em et~al.\ }\relax, }2005]{durr2005role}
D{\"u}rr, Detlef, Goldstein, Sheldon, Tumulka, Roderich, \& Zangh\`i, Nino.
  2005.
\newblock On the role of density matrices in Bohmian mechanics.
\newblock {\em Foundations of Physics}, {\bf 35}(3), 449--467.

\bibitem[\protect\citename{Earman, }2006]{earman2006past}
Earman, John. 2006.
\newblock The ``past hypothesis'': Not even false.
\newblock {\em Studies in History and Philosophy of Science Part B: Studies in
  History and Philosophy of Modern Physics}, {\bf 37}(3), 399--430.

\bibitem[\protect\citename{Eddington, }1928]{EddingtonNPW}
Eddington, Arthur~Stanley. 1928.
\newblock {\em The Nature of the Physical World}.
\newblock New York: Macmillan.

\bibitem[\protect\citename{Feynman, }2017]{feynman2017character}
Feynman, Richard. 2017.
\newblock {\em The Character of Physical Law}.
\newblock Cambridge: MIT press.

\bibitem[\protect\citename{Forrest, }1988]{forrest1988quantum}
Forrest, Peter. 1988.
\newblock {\em Quantum metaphysics}.
\newblock Blackwell Publisher.

\bibitem[\protect\citename{Frigg, }2007]{frigg2007field}
Frigg, Roman. 2007.
\newblock A field guide to recent work on the foundations of thermodynamics and
  statistical mechanics.
\newblock {\em The Ashgate companion to the new philosophy of physics},
  99--196.

\bibitem[\protect\citename{Frisch, }2005]{frisch2005counterfactuals}
Frisch, Mathias. 2005.
\newblock Counterfactuals and the past hypothesis.
\newblock {\em Philosophy of Science}, {\bf 72}(5), 739--750.

\bibitem[\protect\citename{Frisch, }2007]{frischcausation}
Frisch, Mathias. 2007.
\newblock Causation, counterfactuals, and entropy.
\newblock {\em In:} Price, Huw, \& Corry, Richard (eds), {\em Causation,
  physics, and the constitution of reality: Russell's republic revisited}.
\newblock Oxford University Press.

\bibitem[\protect\citename{Goldstein, }2001]{goldstein2001boltzmann}
Goldstein, Sheldon. 2001.
\newblock Boltzmann's approach to statistical mechanics.
\newblock {\em Pages  39--54 of:} Bricmont, J., D\"urr, D., Galavotti, M.~C.,
  Ghirardi, G., Petruccione, F., \& Zangh\`i, N. (eds), {\em Chance in
  Physics}.
\newblock Berlin: Springer.

\bibitem[\protect\citename{Goldstein, }2012]{goldstein2012typicality}
Goldstein, Sheldon. 2012.
\newblock Typicality and notions of probability in physics.
\newblock {\em Pages  59--71 of:} {\em Probability in physics}.
\newblock Springer.

\bibitem[\protect\citename{Goldstein \& Teufel, }2001]{goldstein2001quantum}
Goldstein, Sheldon, \& Teufel, Stefan. 2001.
\newblock Quantum spacetime without observers: ontological clarity and the
  conceptual foundations of quantum gravity.
\newblock {\em Physics meets Philosophy at the Planck scale},  275--289.

\bibitem[\protect\citename{Goldstein \& Tumulka, }2011]{goldstein2010approachB}
Goldstein, Sheldon, \& Tumulka, Roderich. 2011.
\newblock Approach to thermal equilibrium of macroscopic quantum systems.
\newblock {\em Pages  155--163 of:} {\em Non-Equilibrium Statistical Physics
  Today: Proceedings of the 11th Granada Seminar on Computational and
  Statistical Physics, AIP Conference Proceedings},  vol. 1332.
\newblock American Institute of Physics, New York.

\bibitem[\protect\citename{Goldstein \& Zangh\`i, }2013]{goldstein2013reality}
Goldstein, Sheldon, \& Zangh\`i, Nino. 2013.
\newblock Reality and the role of the wave function in quantum theory.
\newblock {\em The wave function: Essays on the metaphysics of quantum
  mechanics},  91--109.

\bibitem[\protect\citename{Goldstein {\em et~al.\ }\relax,
  }2010a]{goldstein2010approach}
Goldstein, Sheldon, Lebowitz, Joel~L, Mastrodonato, Christian, Tumulka,
  Roderich, \& Zangh\`i, Nino. 2010a.
\newblock Approach to thermal equilibrium of macroscopic quantum systems.
\newblock {\em Physical Review E}, {\bf 81}(1), 011109.

\bibitem[\protect\citename{Goldstein {\em et~al.\ }\relax,
  }2010b]{goldstein2010long}
Goldstein, Sheldon, Lebowitz, Joel~L, Tumulka, Roderich, \& Zangh{\`\i}, Nino.
  2010b.
\newblock Long-time behavior of macroscopic quantum systems: Commentary
  accompanying the English translation of {J}ohn von {N}eumann's 1929 article
  on the quantum ergodic theorem.
\newblock {\em The European Physical Journal H}, {\bf 35}(2), 173--200.

\bibitem[\protect\citename{Goldstein {\em et~al.\ }\relax,
  }2020]{goldstein2019gibbs}
Goldstein, Sheldon, Lebowitz, Joel~L, Tumulka, Roderich, \& Zangh\`i, Nino.
  2020.
\newblock {G}ibbs and {B}oltzmann entropy in classical and quantum mechanics.
\newblock {\em In:} Allori, Valid (ed), {\em Statistical Mechanics and
  Scientific Explanation: Determinism, Indeterminism and Laws of Nature}.
\newblock Singapore: World Scientific.

\bibitem[\protect\citename{Hicks \& van Elswyk, }2015]{hicks2015humean}
Hicks, Michael~Townsen, \& van Elswyk, Peter. 2015.
\newblock Humean laws and circular explanation.
\newblock {\em Philosophical Studies}, {\bf 172}(2), 433--443.

\bibitem[\protect\citename{Hildebrand, }2013]{hildebrand2013can}
Hildebrand, Tyler. 2013.
\newblock Can primitive laws explain?
\newblock {\em Philosophers' Imprint}.

\bibitem[\protect\citename{Horwich, }1987]{horwich1987asymmetries}
Horwich, Paul. 1987.
\newblock {\em Asymmetries in Time: Problems in the Philosophy of Sciences}.
\newblock MIT Press.

\bibitem[\protect\citename{Hubert \& Romano, }2018]{Hubert2018}
Hubert, Mario, \& Romano, Davide. 2018.
\newblock The wave-function as a multi-field.
\newblock {\em European Journal for Philosophy of Science}, {\bf 8}(3),
  521--537.

\bibitem[\protect\citename{Lanford, }1975]{lanford1975time}
Lanford, Oscar~E. 1975.
\newblock Time evolution of large classical systems.
\newblock {\em Pages  1--111 of:} Moser, J (ed), {\em Dynamical systems, theory
  and applications}.
\newblock Springer.

\bibitem[\protect\citename{Lazarovici \& Reichert,
  }2015]{lazarovici2015typicality}
Lazarovici, Dustin, \& Reichert, Paula. 2015.
\newblock Typicality, irreversibility and the status of macroscopic laws.
\newblock {\em Erkenntnis}, {\bf 80}(4), 689--716.

\bibitem[\protect\citename{Lebowitz, }2008]{lebowitz2008time}
Lebowitz, Joel~L. 2008.
\newblock Time's arrow and {B}oltzmann's entropy.
\newblock {\em Scholarpedia}, {\bf 3}(4), 3448.

\bibitem[\protect\citename{Lewis, }1979]{LewisCDTA}
Lewis, David. 1979.
\newblock Counterfactual Dependence and Time's Arrow.
\newblock {\em No\^{u}s}, {\bf 13}, 455--76.

\bibitem[\protect\citename{Lewis, }1983]{LewisNWTU}
Lewis, David. 1983.
\newblock New Work for a Theory of Universals.
\newblock {\em Australasian Journal of Philosophy}, {\bf 61}, 343--77.

\bibitem[\protect\citename{Loewer, }2001]{loewer2001determinism}
Loewer, Barry. 2001.
\newblock Determinism and chance.
\newblock {\em Studies in History and Philosophy of Science Part B: Studies in
  History and Philosophy of Modern Physics}, {\bf 32}(4), 609--620.

\bibitem[\protect\citename{Loewer, }2007]{LoewerCatSLaw}
Loewer, Barry. 2007.
\newblock Counterfactuals and the Second Law.
\newblock {\em In:} Price, Huw, \& Corry, Richard (eds), {\em Causation,
  Physics, and the Constitution of Reality: Russell's Republic Revisited}.
\newblock Oxford University Press.

\bibitem[\protect\citename{Loewer, }2012]{loewer2012two}
Loewer, Barry. 2012.
\newblock Two accounts of laws and time.
\newblock {\em Philosophical Studies}, {\bf 160}(1), 115--137.

\bibitem[\protect\citename{Loewer, }2020]{loewer2016mentaculus}
Loewer, Barry. 2020.
\newblock The {M}entaculus.
\newblock {\em In:} Barry~Loewer, Brad~Weslake, Eric~Winsberg (ed), {\em Time's
  Arrows and the Probability Structure of the world}.
\newblock Harvard University Press, forthcoming.

\bibitem[\protect\citename{Maudlin, }2007]{MaudlinMWP}
Maudlin, Tim. 2007.
\newblock {\em The Metaphysics Within Physics}.
\newblock New York: Oxford University Press.

\bibitem[\protect\citename{North, }2011]{north2011time}
North, Jill. 2011.
\newblock Time in thermodynamics.
\newblock {\em The oxford handbook of philosophy of time},  312--350.

\bibitem[\protect\citename{Penrose, }1979]{penrose1979singularities}
Penrose, Roger. 1979.
\newblock Singularities and Time-Asymmetry.
\newblock {\em Pages  581--638 of:} Hawking, SW, \& Israel, W (eds), {\em
  General relativity}.
\newblock Cambridge: Cambridge University Press.

\bibitem[\protect\citename{Penrose, }1989]{roger1989emperor}
Penrose, Roger. 1989.
\newblock {\em The Emperor's New Mind: Concerning Computers, Minds, and the
  Laws of physics}.
\newblock Oxford: Oxford University Press.

\bibitem[\protect\citename{Price, }2004]{price2004origins}
Price, Huw. 2004.
\newblock On the origins of the arrow of time: Why there is still a puzzle
  about the low-entropy past.
\newblock {\em Contemporary debates in philosophy of science},  219--239.

\bibitem[\protect\citename{Reichenbach, }1956]{reichenbach1956direction}
Reichenbach, Hans. 1956.
\newblock {\em The direction of time}.
\newblock  Vol. 65.
\newblock Univ of California Press.

\bibitem[\protect\citename{Rovelli, }2020]{rovelli2020memory}
Rovelli, Carlo. 2020.
\newblock Memory and entropy.
\newblock {\em arXiv preprint arXiv:2003.06687}.

\bibitem[\protect\citename{Von~Neumann, }1955]{von1955mathematical}
Von~Neumann, John. 1955.
\newblock {\em Mathematical foundations of quantum mechanics}.
\newblock Princeton University Press.

\bibitem[\protect\citename{Wallace, }2012]{wallace2012emergent}
Wallace, David. 2012.
\newblock {\em The Emergent Multiverse: Quantum theory according to the Everett
  interpretation}.
\newblock Oxford: Oxford University Press.

\bibitem[\protect\citename{Wallace \& Timpson, }2010]{wallace2010quantum}
Wallace, David, \& Timpson, Christopher~G. 2010.
\newblock Quantum mechanics on spacetime {I}: Spacetime state realism.
\newblock {\em The British Journal for the Philosophy of Science}, {\bf 61}(4),
  697--727.

\bibitem[\protect\citename{Wilhelm, }2019]{wilhelm2019typical}
Wilhelm, Isaac. 2019.
\newblock Typical: A Theory of Typicality and Typicality Explanations.
\newblock {\em The British Journal for the Philosophy of Science}.

\bibitem[\protect\citename{Winsberg, }2004]{winsberg2004can}
Winsberg, Eric. 2004.
\newblock Can Conditioning on the ``Past Hypothesis'' Militate Against the
  Reversibility Objections?
\newblock {\em Philosophy of Science}, {\bf 71}(4), 489--504.

\end{thebibliography}


\end{document}